\documentclass[10pt,journal,compsoc]{IEEEtran}
\usepackage{amsmath,amssymb,amsfonts}
\usepackage{bbding}
\usepackage{tabularx}
\usepackage[printwatermark]{xwatermark}
\usepackage{xcolor}
\usepackage{graphicx}

\ifCLASSOPTIONcompsoc
  \usepackage[nocompress]{cite}
\else
  \usepackage{cite}
\fi

\usepackage{titlesec}

\titlespacing\section{0pt}{10pt plus 2pt minus 2pt}{0pt plus 2pt minus 2pt}
\titlespacing\subsection{0pt}{10pt plus 2pt minus 2pt}{0pt plus 2pt minus 2pt}
\titlespacing\subsubsection{0pt}{10pt plus 2pt minus 2pt}{0pt plus 2pt minus 2pt}

\usepackage{color,colortbl}
\definecolor{Gray}{gray}{0.9}

\begin{document}
\title{DIMY: Enabling Privacy-preserving \\Contact Tracing}

\author{Nadeem Ahmed, Regio A. Michelin,
        Wanli Xue,~\IEEEmembership{Member,~IEEE,}\\
        Guntur Dharma Putra,~\IEEEmembership{Student Member,~IEEE,}
        Sushmita Ruj,~\IEEEmembership{Senior Member,~IEEE,}\\
        Salil S. Kanhere,~\IEEEmembership{Senior Member,~IEEE,}
        and~Sanjay Jha,~\IEEEmembership{Senior Member,~IEEE,}
        
% \author{Nadeem Ahmed,~\IEEEmembership{Member,~IEEE,}
%         Regio A. Michelin,~\IEEEmembership{Fellow,~OSA,}
%         Wanli Xue,~\IEEEmembership{Fellow,~OSA,}
%         Guntur Dharma Putra,~\IEEEmembership{Fellow,~OSA,}
%         Sushmita Ruj,~\IEEEmembership{Fellow,~OSA,}
%         Salil S. Kanhere,~\IEEEmembership{Fellow,~OSA,}
%         and~Sanjay Jha,~\IEEEmembership{Life~Fellow,~IEEE}% <-this % stops a space
\IEEEcompsocitemizethanks{\IEEEcompsocthanksitem N. Ahmed, R. Michelin, W. Xue, G. Putra, S. Kanhere and S. Jha are with the Cyber Security Cooperative Research Centre (CSCRC) - Australia and University of New South Wales (UNSW) - Sydney, Australia.\protect\\
Corresponding author:  nadeem.ahmed@cybersecuritycrc.org.au
\IEEEcompsocthanksitem S. Ruj is with CSIRO, Data61 - Sydney, Australia.}% <-this % stops an unwanted space
\thanks{Manuscript received November 18, 2020; revised January 26, 2021.}}


\markboth{Journal of \LaTeX\ Class Files,~Vol.~14, No.~8, August~2015}%
{Ahmed \MakeLowercase{\textit{et al.}}: DIMY: Enabling Privacy-preserving Contact Tracing}



\IEEEtitleabstractindextext{%
\begin{abstract}
The infection rate of COVID-19 and lack of an approved vaccine has forced governments and health authorities to adopt lockdowns, increased testing, and contact tracing to reduce the spread of the virus. Digital contact tracing has become a supplement to the traditional manual contact tracing process. However, although there have been a number of digital contact tracing apps proposed and deployed, these have not been widely adopted owing to apprehensions surrounding privacy and security. In this paper, we propose a blockchain-based privacy-preserving contact tracing protocol, "Did I Meet You" (DIMY), that provides full-lifecycle data privacy protection on the devices themselves as well as on the back-end servers, to address most of the privacy concerns associated with existing protocols. We have employed Bloom filters to provide efficient privacy-preserving storage, and have used the Diffie-Hellman key exchange for secret sharing among the participants. We show that DIMY provides resilience against many well known attacks while introducing negligible overheads. DIMY's footprint on the storage space of clients' devices and back-end servers is also significantly lower than other similar state of the art apps.
\end{abstract}

\begin{IEEEkeywords}
COVID-19, Contact Tracing, Bloom Filter, Blockchain, Privacy, Security.
\end{IEEEkeywords}}

\maketitle

\IEEEdisplaynontitleabstractindextext

% \fi
\IEEEpeerreviewmaketitle

\IEEEraisesectionheading{\section{Introduction}\label{sec:introduction}}

%\IEEEPARstart{T}{his} demo file is intended to serve as a ``starter file''
%for IEEE Computer Society journal papers produced under \LaTeX\ using
%IEEEtran.cls version 1.8b and later.
%I wish you the best of success.

 

\IEEEPARstart{T}{he} outbreak of the COVID-19 pandemic has changed many aspects of everyone's way of life. One of the characteristics of COVID-19 is its airborne transmission, which makes it highly contagious. Moreover, a person infected with COVID-19 can be asymptomatic, thus spreading the virus without showing any symptoms. Anyone who comes into close contact\footnote{According to the Centres for Disease Control and Prevention (CDC, https://www.cdc.gov/coronavirus/2019-ncov/downloads/2019-ncov-factsheet.pdf), close contact with an infected person is defined as a contact within a range of 6 feet for approximately 15 minutes.} with an infected person is at a high risk of contracting the coronavirus. The lack of an approved vaccine has led governments to enforce lockdowns, quarantines and to recommend social distancing, aiming to prevent the spread of COVID-19. However, despite these precautionary measures, the rate of spread of COVID-19 is putting the public health systems of many countries under strain.   

Contact tracing %any citation (maybe we can find some examples where contact tracing was used for Ebola and SARS)
is an established technique that has proven successful in dealing with other outbreaks such as Ebola and SARS. Contact tracing aims to establish the close-contacts of an infected person so that they may be tested/isolated to break the chain of infection. Traditionally, the contact tracing process is performed manually in a \emph{reactive} manner, triggered when a person tests positive to the virus. This is achieved by conducting a face-to-face interview to establish contacts made by the person while infectious\footnote{The infectious period for a COVID-19 positive case is considered as 2-3 weeks including the asymptomatic period.}. This approach, although useful, has some limitations: \textit{i}) It requires a large trained workforce to cope with the caseload. \textit{ii}) It is hard for people to remember everyone they have met while infected in the last 2-3 weeks. \textit{iii}) A person may have met people that are strangers. Proactive contact tracing \cite{blueTrace, pact-ec, Apple, Google} has recently been proposed to mitigate these issues by maintaining a record of all close contacts made by a person and utilising these records if that person tests positive.  
 
One way of implementing proactive contact tracing is to mandate record-keeping of people's attendance at public venues such as offices,  restaurants, etc. This can be done manually , for example, through QR codes that direct attendees to register their details. However, this increases the risk to individuals' data privacy and allows for possible tracking of the user's behaviour. A more popular option is to employ smartphone-based digital contact tracing apps that can exchange Bluetooth Low Energy (BLE) messages with each other to record this contact. 
% add some more reference for the BLE
 %The most common tracing used is applied manually, in which the people have to sign up manually before joining to a specific venue. After this procedure, if someone from that venue is identified with COVID-19, using the tracing, everyone that was exposed can be notified. This process presents many different flaws, as it relies upon that the person responsible for the venue will maintain the attends plain data, the staff from the venue will need manually to contact the people from the list. Among the main risks, we can highlight the data leakage and, due to human error, a person could not be notified of the exposure. 
%Due to the risks presented in the manual process, many entities (either governments, companies and researchers) are developing automatic tracing applications, that can identify and notify people of possible exposure to the virus. Most of the proposed tracing application rely on mobile phone utilisation (they assume that everyone carry an smartphone) and are running some application that indicates a possible virus exposition. 
The digital contact tracing app is typically composed of two main entities, the smartphones acting as clients and a back-end server. In this model, the smartphones of two individuals with tracing apps installed would exchange some random identification code (this identification code does not reveal any sensitive information about their actual identities) when they are in close proximity. The back-end is typically maintained by health organisations (or the government), and once a person is diagnosed with COVID-19, they can opt to share the local list of contacts stored on their smartphone with the back-end server to identify at-risk users.

The popularity of digital contact tracing apps can be gauged by the fact that more than 45 such apps have been proposed or are being used in by different countries \cite{mit}. 
%%discuss their main architectures briefly
These COVID-19 digital contact tracing apps are based on different architectures distinguishable in
several aspects, including anonymous ID generation and exchange, risk analysis and notifications, etc. For details, readers are referred to a recent survey on digital contact tracing apps by Ahmed \textit{et al.}~\cite{Ahmed:2020}, in which the architectures are classified as centralised, decentralised and hybrid, according to the distribution of key functionalities among the clients and the back-end server. 
%%introduce the apprehensions around security privacy issues associated with DCT
%1. Privacy issues with Centralised servers collecting data
%2. deanonymisation by server
%3. deanonymising by other users
%4. 
However, recent security and privacy analyses of these apps has revealed several risks and issues \cite{Ahmed:2020}, \cite{V2020}, \cite{V2020a}. These apps operate on different trust models. Apps based on the centralised architecture (such as \cite{opentrace}, \cite{covidsafe}, etc.) generally collect sensitive data at the server, that are assumed to be trusted, and only provide privacy protection against malicious users. This trust model makes these apps vulnerable to server-side breaches and malicious actions by the server. On the other hand, apps based on decentralised and hybrid architectures assume an honest-but-curious server model whereby the server will try to harvest sensitive information, if available. Apps such as \cite{DP3T} and \cite{pact-ec} that are based on the decentralised architecture share the anonymous identifiers of the positive cases with all users for matching, making these apps vulnerable to linkage attacks, whereby malicious users can discover the real identities of persons diagnosed with COVID-19 \cite{V2020}. Apps based on hybrid systems perform the risk analysis and notification process at the server instead of revealing the anonymous IDs of positive cases to other users for matching, as proposed in the decentralised architecture. However, these apps suffer from high communication and processing costs. For example, the DESIRE protocol \cite{Desire} uses three BLE messages to advertise a single anonymous ID from a device \cite{DP3T}, while the ContraCorona app \cite{contra} employs three non-colluding servers (submission, matching and notification servers) to manage the contact tracing process.%and maintains two encounter identifiers lists at each device.
% We need to highlight at least one issue in the hybrid solution. Possible issues... lot of communication, proccessing.. .we need to check and identify it.

In this paper, we propose a new privacy-preserving digital contact tracing protocol called "Did I Meet You" (DIMY) that can be classified in the hybrid category.  We take a holistic view of the privacy and security requirements for digital contact tracing and employ techniques to address most of the concerns associated with existing contact tracing protocols.  % Does BlockChain qualify as Honest-but-curious or any other attacker model
We make the following specific contributions:
\begin{itemize}
\item DIMY has been designed to provide full life cycle data privacy protection that prevents contact tracing data from being used arbitrarily. This is achieved by using the Diffie-Hellman key exchange and a secret sharing mechanism, to establish a secret contact representation between user devices over an inherently insecure BLE broadcast channel. We also employ Bloom Filters for storing close contact information both at the individual device level as well as the back-end. Additionally, information from multiple close contacts are stored in a single fixed-size Bloom filter. This contact information is deleted from the user's device once it is encoded in the Bloom filter, which serves two important purposes: (i) It prevents information leakage not only at the client level (for example as a result of device theft or coercion attacks), but also from authorities operating the back-end and governments that can obtain subpoenas to access information stored on the back-end. (ii) It considerably reduces client device and back-end storage requirements.  
\item As opposed to traditional apps that employ centralised servers at the back-end, we have improved the scalability and security of our proposed solution by employing a blockchain-based back-end design in the ecosystem. %for storage with the smart contract to perform the exposure risk analysis. 
This provides transparency and trust on back-end operations besides ensuring the integrity of data uploads from positively identified cases that are appended as blockchain transactions. We also evaluate the performance of our implementation based on  Hyperledger and show that DIMY provides low latency and resource consumption while supporting high throughput under moderate loads.
\item We consider a comprehensive threat model and provide safeguards against several types of adversaries including malicious users, back-end (admin and developers) and government (discussed in detail in Section \ref{subsec:threat}). We also provide a comprehensive security and privacy analysis of our proposed solution and show that DIMY provides resilience against common attacks such as linkage, enumeration, social graph construction and replay (discussed in detail in Section \ref{subsec:resilience}). 
\end{itemize}

This paper is organised as follows. We discuss related work in Section \ref{sec:rw}. Section \ref{sec:Back} introduces the background information necessary to understand the building blocks of our proposed solution. We detail the design of our DIMY protocol in Section \ref{sec:Protocol}. In Section \ref{sec:Comp}, we compare the salient features of DIMY with other existing protocols. Section \ref{sec:s&p} provides a security and privacy analysis of DIMY protocol, while Section \ref{sec:evaluation} details the performance analysis of our proposed solution. Section \ref{sec:Conc} concludes this paper.

\section{Related work}
\label{sec:rw}
There have been a number of digital contact tracing apps proposed, developed and deployed to aid in identifying exposure from infected individuals. Most of the apps are based on BLE message exchanges, while some of the proposed apps also employ location tracking based on GPS. The MIT Technology Review summarised the salient features of 47 such apps \cite{mit}. These apps follow different approaches for development and addressing multiple aspects in terms of privacy, security, performance, and reliability etc.  We follow the classification criteria discussed by Ahmed et al., in \cite{Ahmed:2020} to classify tracing apps in centralised, decentralised and hybrid categories according to the underlying application architecture and the functionalities delegated to client devices and the server. 

\subsection {Centralised approach}
BlueTrace protocol \cite{blueTrace} is one of the first proposed digital contact tracing protocols that is based on a centralised architecture. This protocol was employed to develop the Singaporean TraceTogether \cite{opentrace} and the Australian CovidSafe \cite{covidsafe} apps. %The main implementation of a centralised architecture is given through the BlueTrace protocol implementation, which was one of the first contact tracing protocols. The TraceTogether app is one of the first apps released in March 2020 by the Singaporean government which follows the BlueTrace protocol specification. In April 2020, Australian government released the CovidSafe (AU) [] app also following the BlueTrace protocol. In the other hand, Aarogya Setu [] tracing app was developed by Indian government, and does not follow the BlueTrace protocol, but sill relies in a centralised architecture. 
Another protocol named ROBERT \cite{robert} was proposed by Inria and Fraunhofer AIESEC that is also based on the centralised architecture.% which focus in the privacy-preserving on the contact tracing.

In the centralised architecture, a central server is responsible for handling major components of the digital contact tracing process such as ID generation, risk analysis and notification, etc. Typically in this architecture, a user enrols with the central authority, which periodically (typically every 10-15 min) generates a unique temporary ID for each client. This temporary ID is sent to the user and is used in his/her advertisement message. The user records the received temporary IDs locally when in the proximity of other contacts running the same app.
%that it receives following contact with another user that has the app installed. This encounter is recorded locally in the mobile phone of both users. 
If a user gets diagnosed with COVID-19, a health officer authorises the user to upload (share) the list of all captured IDs to the centralised server for risk analysis and notification of close contacts. 

%One characteristic that is different in these apps is related to the generated ID lifetime, for example, while TraceTogether lifetime of each ID is 15 minutes, CovidSafe uses a 2 hours temporary ID. The protocol ROBERT, defines a time $M$ (which allows the app developer specify it) that is the validity of user identification, once it expires, the client need to query the server to a new set of Ephemeral ID. %In the other hand Aarogya Setu specification, does not presents any information related refreshing or recreating the user identification.

The central server in the BlueTrace protocol can access the personally identifiable information collected at the registration stage and map each client to their temporary IDs. This raises issues with privacy as this sensitive data can be used for other purposes besides digital contact tracing.
In contrast to BlueTrace, ROBERT protocol does not store any user identifiable information on the server. Temporary IDs are still created at the server without been linked with the devices used by the clients. %Once diagnosed, a user will upload all of the contact IDs collected in a specified period. 
ROBERTS's notification process requires the uploading of IDs used by a device to check whether they have come in contact with a COVID-19 positive case or not. This is in contrast with BlueTrace where the server can identify at-risk users and contact them proactively. 

ROBERT protocol, however, similar to other protocols based on centralised architectures, has a high potential to \emph{function creep}, in which it can be re-purposed into a mass surveillance system \cite{DP3T}. Another potential issue associated with centralised architectures is the construction of partial social graphs (discussed in detail in Section \ref{sec:s&p}) that enable linkability of infected cases and their contacts. A server breach can also result in the loss of sensitive data stored at the server.    

%Due to the centralisation of processing in the server, it requires some information from the clients. Each protocol and app can share a different set of information with the server. BlueTrace protocol typically requests from the user the name, mobile phone number, age and post code upon the registration. Next the encounters tuples are composed by the TempID, phone model and transmit power which are stored locally in the users mobile phone. These information are shared with the server only upon receiving an authorisation from the Health Authority. Aarogya Setu requests name, age, occupation and foreign travel history in the last 30 days, for the registration process. The encounter combines information retrieved from Bluetooth and GPS data, and this information is shared with the server. The data is stored for 30 days before deleting it. ROBERT protocol does not define any personal information request upon registration, in fact, in this step the client share a key, the epoch time, duration of an epoch,  with the server. The server will generate a permanent identifier and associate with a list of ephemeral identifiers. The client will exchange hello messages with the neighbours and once receive an authorisation from the Health Authority, the user shares with the server a list of Hello and Time when the encounter occurred.

%Robert: https://hal.inria.fr/hal-02611265/document

% \begin{itemize}
%     \item IDs type-generation-lifetime
% \item What is uploaded
% \item How the Risk analysis/notification is performed and by whom
% \item Issues (privacy/security attacks)
% \end{itemize}

\subsection{Decentralised approach} 
The decentralised architecture differs from the centralised version by pushing some functionalities to the user's devices. There is still a server involved, however, the role played by the server is more in terms of orchestrating the clients. This approach claims to improve user privacy by generating temporary IDs in the user's devices. Additionally, exposure risk processing is also performed at the device level. %Among the tracing app that are implemented following the decentralised architecture, we highlight PACT-East[] and DP-3T[].

Generally, devices generate random seeds for forming their temporary IDs. These IDs are exchanged with other users who they come in contact with. Once a user is diagnosed positive with COVID-19, all seeds used by the device (some of the apps upload IDs instead of seeds) are uploaded to the server. Any user who wishes to check whether they are at-risk can download the seeds (or IDs) uploaded by the diagnosed users. The device can then perform matching locally, with the user notified of the result. The server is neither involved in the ID generation nor the at-risk analysis and notification process. 

There are a number of protocols that follow the decentralised architecture such as DP-3T \cite{DP3T}, PACT-East Coast \cite{pact-ec}, Google Apple Encounter Notification (GEAN) \cite{Google}, \cite{Apple} and TCN \cite{TCN}. They have minor differences in the implementation of sub-components with the basic design following the general functionality described in this section.

Apps based on decentralised architectures provide enhanced privacy protection against server-based attacks as devices generate their own anonymous IDs. However, decentralised apps are known to be vulnerable to linkability attacks, whereby a user who has received the IDs generated by an infected user is able to link the IDs with the real user's identity \cite{V2020}. These apps are also subject to enumeration attacks, enabling the counting of all positive cases by each user. 
%The temporary identification, in the decentralised architecture, is managed in the user's device. Following this approach, the lifetime of each identification. 

%DP-3T [] introduces the k-out-of-n secrete sharing to improve the user's privacy and protected the identity when exchanging credentials in a close contact. Additionally, the server in this protocol is responsible for  converting the temporary identification received (Ephemeral IDs) in a Cuckoo filter to identify possible exposures positive cases. The data is stored locally in each user's device for 14 days. 

%Similarly to DP-3T, DIMY uses the K-out-of-N (Shamir Secret Sharing) on the user's device to share the keys, which means that even if a malicious user intercept a single package, it will not disclosure any relevant information from its source. Another difference between DP-3T and DIMY, is while the former upload the contacts using a Cuckoo filter, the later uses Bloom filter to upload the contacts.

% \begin{itemize}
%     \item IDs type-generation-lifetime

% \item What is uploaded:
% Approach A: list of your IDs for incubation period
% Approach B: list of all IDs captured for the incubation period
% \item How the Risk analysis/notification is performed and by whom
% \item Issues (privacy/security attacks)
% \end{itemize}

\subsection{Hybrid approach}
Hybrid architectures balance the tasks between the client and the server. %According to \cite{Ahmed:2020}, in the hybrid architecture, 
The server is responsible for performing the risk analysis and notification process, while the client manages the generation of temporary IDs. Desire~\cite{Desire} is one of the example protocols that follow the hybrid architecture. 
Devices using the hybrid protocol cryptographically generate and exchange IDs with other devices. A contact between two devices is represented by a unique encounter ID, which the app generates by combining own and the received temporal IDs. A user who tests positive can optionally upload the generated encounter IDs to the server. Any user who wants to check the risk of exposure sends their collected encounter IDs to the server for matching. The server performs risk analysis and notifies any user who is deemed to be at risk.  

The Desire protocol uses 256-bit IDs that are broadcast for generating encounter IDs (or tokens) using the Diffie-Hellman key exchange. This design choice requires at least three BLE message exchanges (Advertisement, Scan\_Request and Scan\_Response) resulting in an increase in energy consumption for devices~\cite{DP3T2}. %Moreover, the encounter IDs are stored at the server for matching with requests from other users. This may result in leakage of the number of contacts for infected users, especially if combined with other side-channel information.

%Once the app is installed in the user's device, it connected to a server to register. This operation does not exchange any personal information between client and server.  The device is responsible for generate the Ephemeral IDs and manage them. Typically these EphID are valid for 15 minutes.
% \begin{itemize}
%     \item IDs type-generation-lifetime
% \item What is uploaded
% \item How the Risk analysis/notification is performed and by whom
% \item Issues (privacy/security attacks)
% \end{itemize}
%\subsection{Why a new Digital Contact Tracing Protocol?}
\subsection{Discussion}
We have listed the modalities involved in the design of the three types of architectures commonly used for digital contact tracing. We have also highlighted some common issues related to privacy and security that are associated with apps based on these architectures. Our proposed solution, DIMY, can be broadly classified as a hybrid architecture in that we generate the IDs on the devices, and perform risk-analysis and notification tasks at the server. 

For DIMY, we utilise Bloom filters to encode the encounter ID generated by the devices and to store the encounters at the back-end. The DP-3T protocol also suggests the use of Cuckoo filters, but in a different context. In their proposal, the server has access to all the seeds uploaded by users who have tested positive, and hence can generate the IDs used by positive cases. They proposed encoding these IDs in a Cuckoo filter to hide them from other users who are performing local risk-analysis. In comparison, our use of Bloom filters provides better privacy protection as these are used to hide the encounter information both at the device level as well as at the back-end.

%We also note that other tracing apps are also 
We also employ blockchain technology to manage the back-end processing. BeepTrace~\cite{xu2020beeptrace} is another framework that has proposed the use of two blockchains: `tracing chain' to manage tracing/contact matching with anonymised user data, and the other `notification chain' to manage notifications at the back-end.  In contrast, we propose using a single blockchain (Hyperledger Fabric) and enhancing its privacy protection further by using Bloom filter-encoded data storage. Additionally, we rely on the smart contract functionality to perform the exposure risk-analysis and matching in a privacy-preserving manner. 
Lv et. al. \cite{LWJCQZ2020} propose \emph{Bychain}, a new blockchain that stores contact information securely. Bychain protects user identities using Zero-Knowledge protocols. Though the contact information is stored on the blockchain, the authors do not discuss how data is retrieved and used when individuals test positive for COVID-19. In addition, the proposed protocols rely on support from GPS equipped provers and witnesses for recording contact information using LTE, WiFi and BLE. Our proposed design, in contrast, is an end-to-end BLE based solution for contact tracing relying on widely available BLE modules on smartphones.

We also note that in the context of digital contact tracing, use of cryptographically generated IDs and the Diffie-Hellman key exchange mechanism has been proposed in \cite{contra}, \cite{Pronto}, \cite{Desire}. Similarly, the $k$-out-of-$n$ secret sharing mechanism for ID distribution has been proposed as extension to the standard protocols in ~\cite{DP3T}, \cite{contra}. In our proposed protocol, the secret sharing mechanism is coupled with the Diffie-Hellman key exchange. We integrate these security and privacy-preserving techniques with efficient set membership using Bloom filters and additionally employ blockchain technology at the back-end. Table~\ref{table:relatedWorkProtocols} highlights the key technologies used in DIMY and places our proposal in the context of existing protocols.
%%Need to look at BeepTrace and ByChain for classification of architecture
\begin{table}[ht]
\vspace{-4mm}
\centering
\caption{Comparison of key technologies (C=Centralised D=Decentralised H=Hybrid). A $\star$ denotes an extension to the base protocol.}
\vspace{-4mm}
\label{table:relatedWorkProtocols}
\begin{tabular}{l|c|c|c|c|c|}
\cline{2-6}
                                                                                         & \multicolumn{5}{c|}{Key Technologies}                                              \\ \hline
\multicolumn{1}{|l|}{Protocols}                                                          & DH & \begin{tabular}[c]{@{}c@{}}Secret\\ Sharing\end{tabular} & BF & BC & Architecture \\ \hline
\multicolumn{1}{|l|}{BlueTrace{~\cite{blueTrace}}}                                                   & $\times$  & $\times$                                                        & $\times$  & $\times$  & C            \\ \hline
\multicolumn{1}{|l|}{CovidSafe{~\cite{covidsafe}}}                                               & $\times$  & $\times$                                                        & $\times$  & $\times$  & C            \\ \hline
\multicolumn{1}{|l|}{ROBERT{~\cite{robert}}}                                                      & $\times$  & $\times$                                                        & $\times$  & $\times$  & C            \\ \hline
\multicolumn{1}{|l|}{DP-3T{~\cite{DP3T}}}                                                      & $\times$  & $\star$                                                         & \Checkmark  & $\times$  & D            \\ \hline
\multicolumn{1}{|l|}{\begin{tabular}[c]{@{}l@{}}PACT-East \\ Coast{~\cite{pact-ec}}\end{tabular}} & $\times$  & $\times$                                                        & $\times$  & $\times$  & D            \\ \hline
\multicolumn{1}{|l|}{GAEN{~\cite{Apple},~\cite{Google}}}                                                   & $\times$  & $\times$                                                        & $\times$  & $\times$  & D            \\ \hline
%\multicolumn{1}{|l|}{TCN{[}13{]}}                                                        & $\times$  & $\times$                                                        & $\times$  & $\times$  & D            \\ \hline
\multicolumn{1}{|l|}{Desire{~\cite{Desire}}}                                                     & \Checkmark  & $\times$                                                        & $\times$  & $\times$  & H            \\ \hline
\multicolumn{1}{|l|}{Contra Corona{~\cite{contra}}}                                              & \Checkmark  &  $\star$                                                        & $\times$  & $\times$  & H            \\ \hline
\multicolumn{1}{|l|}{BeepTrace{~\cite{xu2020beeptrace}}}                                                  & $\times$  & $\times$                                                        & $\times$  & \Checkmark  & H           \\ \hline
\multicolumn{1}{|l|}{ByChain{~\cite{LWJCQZ2020}}}                                                  & $\times$  & $\times$                                                        & $\times$  & \Checkmark  & H           \\ \hline
\multicolumn{1}{|l|}{DIMY}                                                               & \Checkmark  & \Checkmark                                                        & \Checkmark  & \Checkmark  & H            \\ \hline
\end{tabular}
\vspace{-4mm}
\end{table}

%%TraceSecure%%

%It combines uses an entity responsible for key distribution/management. BeepTrace also introduces a role of a Geo Solver entity responsible for reading the raw data from the blockchain, compute the matches and send the notifications to the blockchain. The authors proposed the development of a new blockchain to support the solution, and they point some limitations related to the mining and incentives mechanisms.

%%%https://arxiv.org/pdf/2005.10103.pdf
%Related work using BC
%Xu et al.~\cite{xu2020beeptrace} introduced the BeepTrace, which is a blockchain based application enabling the user privacy preserving for pandemic scenarios. In Xu's research, the blockchain provides the infrastructure to integrate the patient or user and health authorities.

\section{Background information}
\label{sec:Back}

In this section, we introduce key technologies which form the building blocks of our proposed solution, including Diffie Hellman key exchange, Shamir secret sharing, Bloom filters, and blockchain.

\subsection{Diffie Hellman Key Exchange}
\label{subsec:dh}
%%Nadeem%%
%There are two popular ways to secure communications using secret keys. In public key cryptography, each user generates two keys, an encryption key $K_{Enc}$ and the corresponding decryption key $K_{Decr}$ such that computing $K_{Decr}$ from $K_{Enc}$ is computationally infeasible. A user publicly advertises its $K_{Enc}$ and keeps the $K_{Decr}$ private.
%\textcolor{blue}{Do we need this paragraph?}
Diffie-Hellman \cite{Diffie} is a public key distribution system that addresses the issue of secret key distribution over an insecure channel. It enables two users to communicate with each other in order to arrive at a common symmetric secret key  that can be used for encrypting/decrypting their future communications. This secret key is computed in such a manner that an eavesdropper cannot reconstruct the shared secret key, in a computationally feasible context, even if they have heard all the messages exchanged. 

This key distribution mechanism is based on the discrete logarithm problem. %Both communicating parties use a common generating element $g$ out of a multiplicative group of elements $G$ of order $n$. Each party selects a random number $r$ from space $n$ and computes $g^{r_i}$. The parties now exchange these values with each other. They can now compute the shared symmetric secret where $(g^{r1})^{r2} = (g^{r2})^{r1}$ assuming $r1$ and $r2$ are the random numbers generated by the two parties. Note that $g$, $g^{r1}$ and $g^{r2}$ are known to any third party that can capture the message exchange, but they are unable to compute the shared secret.This key exchange mechanism can also be implemented using Elliptic curves cryptography.
Let $G$ be a multiplicative group of prime order $n$. Let $g \in G$ be a generator. Party $A$ chooses $r_1 \in Z_p$, computes $g^{r_1}$ and sends to party $B$. 
$B$ chooses $r_2 \in Z_p$, computes $g^{r_2}$ and sends to party $A$. 
On receiving $g^{r_2}$, A computes $(g^{r_2})^{r_1} = g^{r_1r_2}$, similarly, 
on receiving $g^{r_1}$, B computes $(g^{r_1})^{r_2} = g^{r_1r_2}$. Due to the hardness of the discrete logarithm problem, an adversary cannot compute $r_1$, given $g^{r_1}$. Hence, it cannot construct the common key. In our contact tracing protocol, $G$ is an elliptic curve group. %\textcolor{blue}{Good to specify which curve we have chosen}. 
%secp128r2
\subsection{Shamir Secret Sharing}
\label{sec:k-out-n}
%%Wanli%%
%%Done%%
%%Cut number of references%%
In our proposed protocol, we use a secret sharing scheme~\cite{sss} to provide information privacy and secure communication between the devices participating in contact sharing. %secure information sharing is being very important a
%,liu2012improved,sun2012authenticated}
%have been proposed. %widely adopted among in various cloud computing related applications. 
The basic idea revolves around making shares of a secret that can be securely distributed over many devices by a threshold secret sharing mechanism. 

A secret sharing scheme consists of two phases, called \textit{sharing} and \textit{reconstruction}. In a $k$-out-of-$n$ secret sharing scheme (also referred to as $(k,n)$-secret sharing scheme), there is a unique player called the dealer who wants to share parts of secret $S$ among $n$ players, $P_1, P_2,...,P_n$. The dealer creates $n$ shares of the secret $S$ ($S_1, S_2,...,S_n$) and sends every player a share (say $S_i$ to player $P_i$) of the secret $S$ in a way that any group of $k$ or more players can reconstruct the secret. All shares are necessary for the reconstruction of the secret if we keep $k = n$.

A $k$-out-of-$n$ secret sharing scheme, in general, has to satisfy the following two properties:
%To be concise with the literature, the main two properties that any secret sharing scheme has to fulfil are:
\begin{enumerate}
    \item Recoverability: The secret can be reconstructed given any $k$ shares.
    
    \item Secrecy: No information can be known about the secret given any number of shares $< k$.
\end{enumerate}
 
A dealer is assumed as honest in standard secret sharing. However, additional information or multiple communication rounds are required to verify the consistency of shares held by various parties leading to the notion of a verifiable secret sharing scheme. 

%Secret sharing is considered to be utilised ideally in the digital contact tracing apps/protocols as the key initialisation/deployment~\cite{Desire}, since it implicitly involves the contact time interval while preventing the target-key get revealed by the attacker easily. 

\subsection{Bloom Filter}
%Wanli%
%%Done%%
%%Can we cut the number of references to one or two%%
We employ Bloom Filter (BF)-based storage for logging contact information on the devices and at the back-end blockchain. A Bloom filter (BF)~\cite{bloom1970space} %,gremillion1982designing, bianchi2012better} 
is a probabilistic data structure used to represent set membership. It supports an efficient mechanism for set membership queries. When queried, the BF will return true (with a false positive) if the queried data exists in the filter. 
A BF is implemented as a bit array $BF[i]$, i $\in (1, n)$, of $n$ bits accessed via $h$ independent hash functions $H_1(x)...H_h(x)$, each of which maps an element $x$ in a set $S$ of $m$ elements %\in \{0,1\}^{*} $ 
to one of the $l$ bits within the bit array. Querying the presence/membership of an element $x$ in the set %\in \{0, 1\}^{*} $ 
using a BF requires checking $ \bigwedge_{j=1}^h BF[H_j (x)] = 1$ (i.e., $\bigwedge_{j=1}^h BF[H_j (x)]$ returns 1 only if all $h$ corresponding bits are set to 1). %Bloom filter-based encoding is widely used as an efficient masking technique for effective computations of string~\cite{durham2012framework,schnell2009privacy,vatsalan2014scalable}, categorical~\cite{lai2006efficient,many2012fast} data and numerical sequence data~\cite{xue2020sequence}.

 %%Check with the other section on FP in BF%%

In BFs, false-positives (FP) are possible, but false-negatives (FN) are not. An FP $ \psi (m, k, n)$ is the probability that a membership test performed for an element $x$ not stored in $BF(S)$ %($S$ refers to the set of elements encoded in the filter) 
returns 1, in which the parameter $m$ specifies the size of the bit array (Bloom filter length), $k$ specifies the number of hash functions, and $n$ is the cardinality of the stored set. Even if an exact expression for $\psi (m, k, n)$ is available~\cite{mitzenmacher2005probability}, virtually all work in the field relies on a simple, but tight, approximation:
\begin{equation}\label{eq:fp}
% \psi (l,h,g) \approx \Big( 1 - [1-1/l]^{gh} \Big)^h \approx \Big( 1- e^{-gh/l} \Big)^h =(1-p)^h ,
\psi (m,k,n) =\Big(1- (1- \frac{1}{m})^{kn}  \Big)^{k}  \approx \Big(   1-e^{-kn  /m } \Big)^ k
\end{equation}

%where $p=e^{-kg/l}$.
Simply, an FP is due to the collision of two different elements being mapped to the same bit position.\\
\vspace{-10pt}
%\noindent \textbf{Bloom filter} has been considered and adopted as a probabilistic data structure for the COVID-19 case, e.g., anonymous contact discovery applications~\cite{canetti2020private,tajan2020approach}. The property of Bloom filter data structure implies its potential to be used as an efficient and privacy-preserving data storage in the COVID-19 contact tracing applications, however, the uncertainty and ``plausible deniability" needs to be justified.

\subsection{Blockchain}\label{subsec:blockchain}
%how the blockchain is used in this protocol, and which property it introduces.
Blockchain technology was initially introduced in 2008 to maintain a public ledger of Bitcoin transactions~\cite{nakamoto2019bitcoin}. This technology allows network participants to create chronologically sequential immutable blocks ensuring integrity, trust and transparency.
%Blockchain technology enables applications design to create decentralised ledger. It was proposed initially to maintain a public ledger of Bitcoin transactions, and
Blockchain technology has since been applied in many different applications, from cryptocurrencies~\cite{nakamoto2019bitcoin,wood2014ethereum} to IoT~\cite{Dedeoglu2020}. It enables the creation of solutions that do not rely on a central authority; rather, the chain is spread over several nodes in a distributed manner. It also ensures information integrity by linking the blocks in the chain through a hash function of the previous blocks.
%%Add different instances especially HL and write something about smart contract as well
%Many different blockchain instances are available either commercial or researches. Given the decentralised and secure design introduced by the blockchain technology, 

There are three main types of blockchain; public, private and permissioned. The public instances are blockchains that allow any peer to participate in the network. Some examples are Bitcoin~\cite{nakamoto2019bitcoin} and Ethereum~\cite{ethereum}. Private blockchains are restricted networks that allow only some nodes to participate while relying on a central authority to manage the nodes. As an example, Ethereum can also run in a private instance, however, while executed privately, there is no connection/interaction with the public instance. In the permissioned blockchain,  a group of participants perform the node access control. The main example is Hyperledger, in which organisations are responsible for managing the network.
%Given the characteristics of the contact tracing protocols, the permissioned blockchain is a more suitable choice to model the solution back-end. We therefore use the Hyperledger Fabric. 
The Hyperledger Fabric~\cite{hyperledger} blockchain instance supports the deployment of chaincode, a small piece of source code developed and embedded in the blockchain, and it is executed once a node sends a transaction to its address. %write in Go or JavaScript languages. 
It uses a consensus based on the Byzantine general's problem, known as RAFT, which defines a leader to conduct an election with the existing nodes connected to a given organisation. This consensus protocol makes the Hyperledger Fabric blockchain a good choice to support our solution, where the organisations are modelled as health authorities (see Section~\ref{sec:Protocol}). Using this blockchain technology in our solution allows for data integrity, transparency of operations and decentralised data storage. 

% why are we presenting the consensus here? how this can be related with our work? leaders ==> health authorities.

\begin{figure*}[!ht]
\centering
\includegraphics[width=0.6\textwidth]{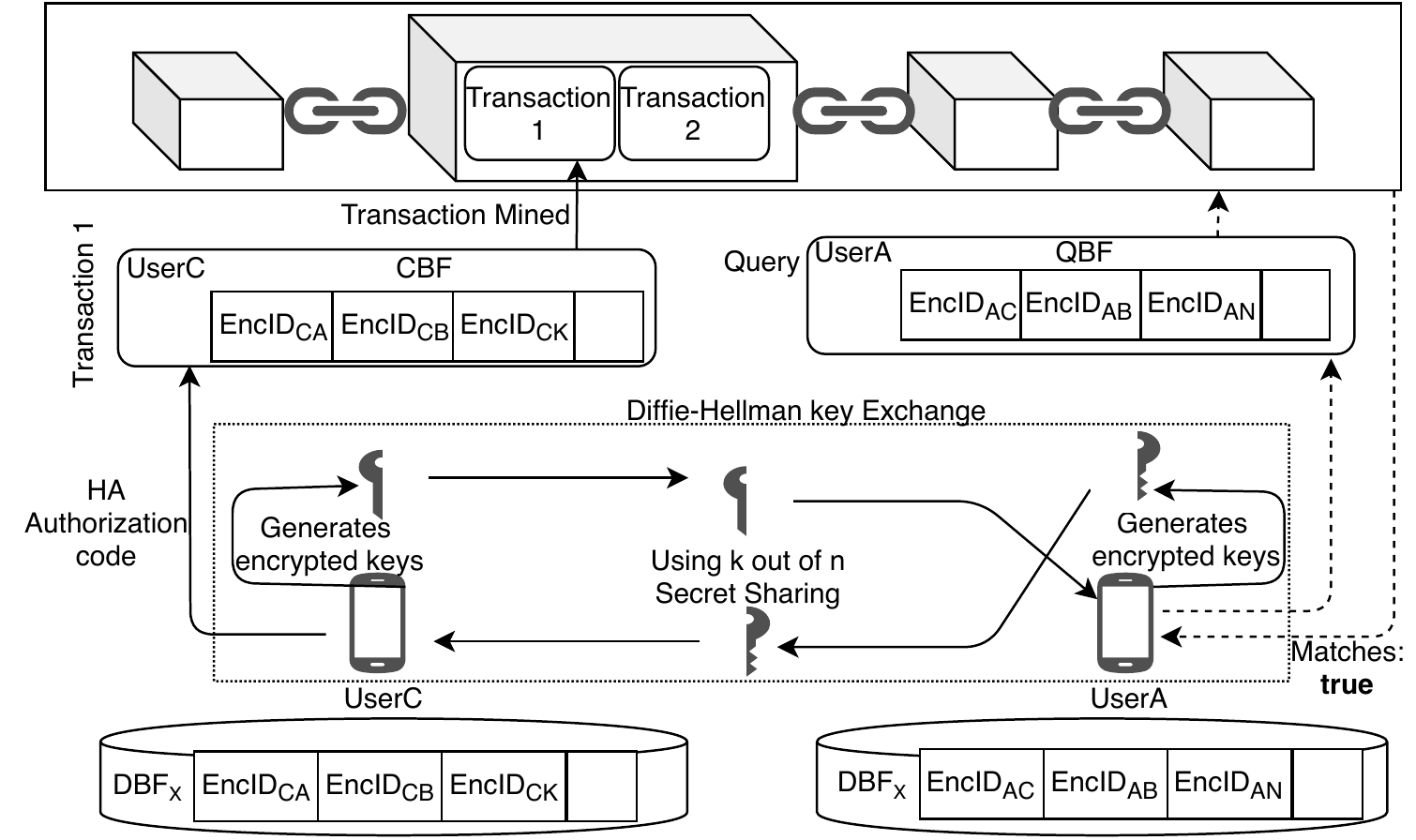}
\caption{Basic protocol architecture.}
\label{fig:platform}
\vspace{-4mm}
\end{figure*}

\section{DIMY Protocol Description}
\label{sec:Protocol}

We first provide an overview of the DIMY protocol with a detailed description of the building blocks appearing in subsequent sections. Figure \ref{fig:platform} shows the overall architecture for our proposed solution. Consistent with other decentralised and hybrid architecture-based contact tracing approaches, devices participating in DIMY periodically generate random ephemeral identifiers. These identifiers are used in the Diffie-Hellman key exchange (Refer to Section \ref{subsec:dh}) to establish a secret key that would represent the encounter between two devices that come in contact with each other. For example, Alice generates a random number $X_{At}$ at time $t$ and calculates its ephemeral identifier $EphID_{At} = g^{X_{At}} \in \{0,1\}^{128}$ ($g \in G$ is a generator and G is an elliptic curve group of order p). After generating their $EphID$, devices employ the $k$-out-of-$n$ secret sharing scheme to produce $n$ secret shares of the $EphIDs$. Devices now broadcast these secret shares, at the rate of one share per minute, through BLE advertisement messages. A device can reconstruct the $EphID$ advertised from another device if it has stayed in the communication range of this device for at least $k$ minutes, enabling it to collect $k$ secret shares of $EphIDs$. Assume that Alice is able to reconstruct the $EphID_{Bt} = g^
{Y_{Bt}}$ advertised by Bob where $Y_{Bt}$ is a random number generated by Bob at time $t$. Alice now computes the secret encounter identifier $EncID_{ABt} = (g^{X_{At}})^{Y_{Bt}}$. Bob also computes the same encounter identifier $Enc_{ABt}$ having received $k$ advertisements from Alice.
%Both Alice and Bob thus store the symmetric secret identifier with RSSI values.
%%Figure showing EphIds, k-out-of-n advertisements, and Encounter IDS

A novel aspect of our proposed solution is the use of Bloom filters for storing contact information. Each device maintains a Daily Bloom Filter (DBF) and inserts all the constructed encounter identifiers in the DBF created for that day. The encounter identifier is deleted as soon as it has been inserted in the Bloom filter. Devices maintain DBF on a 21 days rotation basis, identified as the incubation period for COVID-19. DBFs older than 21 days automatically get deleted. %This means that a device can have a maximum of 21 DBFs at any given instance.

Our solution employs blockchain at the backend. Once a user is diagnosed with COVID-19, they can volunteer to upload their encounter information to the blockchain. Health Authorities (HA) then generate an authorisation access token from the blockchain that is passed on to the device owner. The user's device combines 21 DBF into one Contact Bloom Filter (CBF) and uploads this filter to the blockchain.
The blockchain stores the uploaded CBF as a transaction inside a block (in-chain storage) and appends the block to the chain.

A user who wants to check whether they have come in close contact with any user who was diagnosed positive can query the blockchain on a daily basis. A device combines all of the locally stored DBFs (the maximum number is limited to 21) in a single Bloom filter called the Query Bloom Filter (QBF).
The QBF is part of the query that gets  uploaded to the blockchain. The blockchain matches the QBF with CBF stored as a transaction in the blockchain and returns "matched" or "not matched" as a response. If the response from the blockchain is negative, the device deletes its QBF. Conversely, if the user is found to be at-risk, the QBF is stored separately for further verification by HA in the contact tracing process.

We now explain each component of the proposed solution in more details.

\subsection{Close contact representation}
\label{subsec:CC}
% discussion on whether to use device identifiers or contact identifiers.
In this section, we briefly discuss the notion of encounter representation in the context of contact tracing apps. %The main aim of proactive digital contact tracing is to register the close contacts of two devices such that this contact information can be utilised later to warn a user if one of their close contact tests positive to COVID-19.
One simple way to achieve contact representation involves using device IDs. In this scheme, which we refer to as an ID-based scheme, each device is assigned a temporal ID, either by a central authority server or computed locally at the device. The devices advertise and exchange these IDs. The presence of an ID in the local storage of a device thus represents an encounter with that user (device). In an alternate scheme that we refer to as the shared secret-based scheme, encounters can be represented by a shared secret between two participants. Both participants exchange specific messages to arrive at a shared secret only known to the parties in communication. %Encounter, in this case, is represented by the shared secret. 

\begin{figure}[!h]
\vspace{-4mm}
\centering
\includegraphics[width=0.48\textwidth]{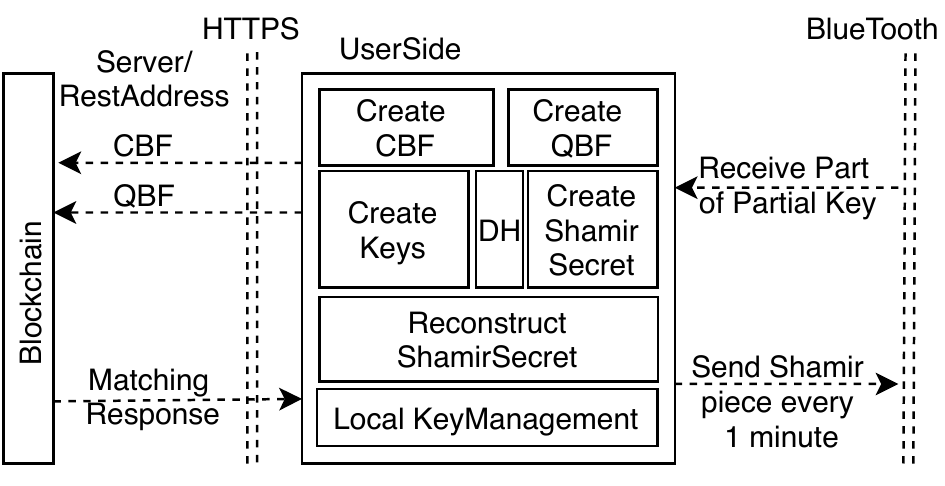}
\vspace{-4mm}
\caption{Information flow in DIMY.}
\label{fig:flow}
\vspace{-4mm}
\end{figure}

In ID-based schemes, all devices in the vicinity of the device A store the same ID advertised by A. In contrast, in the shared secret-based scheme each device pair computes a different shared secret among them. Concretely, if three devices A, B and C meet each other and advertise $ID_A, ID_B$ and $ID_C$ respectively, according to the ID-based scheme, A would store: $\{ID_B,  ID_C\}$, B will store: $\{ID_A, ID_C\}$ and C will store:$\{ID_A, ID_B\}$. If these devices are instead using the shared secret-based scheme, they will end up storing secrets as A: $\{S_{AB},  S_{AC}\}$, B: $\{S_{BA},  S_{BC}\}$ and C: $\{S_{CA},  S_{CB}\}$. 
%where $S_{BC} = S_{CB}$ is the shared secret between devices B and C.
We have used the shared secret-based representation for recording the encounter between neighbouring devices as these provide more resilience against replay attacks discussed in Section \ref{sec:s&p}. We have employed the Diffie-Hellman key exchange scheme for sharing the secret among communication devices. Figure \ref{fig:flow} illustrates the flow of information in the DH scheme used over an insecure BLE communication channel.

%by employing Diffie-Hellman key exchange protocol. 

%There are several advantages of using a shared-secret based scheme as compared to ID-based scheme that we discuss in Section \ref{subsec:cc}. 
%The main reason for this choice would become clear when you discuss how the these encounter identifier are used in the risk analysis and notification process  
%Cannot do replay attacks if the shared secret is being used 

\subsection{Generating identifiers}
\label{subsec:IDs}
%Rewrite the first sentence... it looks consufing.
This component pertains to generating anonymous device IDs that are used as device advertisements. We consider two common design options: i) Each device generates its own pseudo-anonymous random identifier. This is the approach taken by most of the decentralised and hybrid contact tracing apps such as PACT-East Coast, DP-3T and GAEN. ii) A centralised server generates these identifiers for the registered devices that are then periodically transferred to the devices. This approach is used in apps based on a centralised architecture, such as TraceTogether and COVIDSafe (AU).

% Discuss the implications: Server knows about all IDs in the system and can easily identify the contacts. Can build social graphs. 

%Devices in control. Contacts not known unless they explicitly agree and contacts the HO

In our solution, each device generates their ephemeral IDs, which are valid for 30 minutes. This provides privacy protection against exposing a user's contact details (mapping of IDs to real identities) to the back-end. We have kept the size of $EphID$ as 16 Bytes (128 bits), as BLE advertisement messages are only able to carry a limited payload of data. We note that devices do not directly advertise these $EphIDs$; instead, we use the $k$-out-of-$n$ secret sharing mechanism (explained in next section). %We note that current standards suggest using 256 bits group identifiers for using Diffie-Hellman based key exchange. We discuss the implications of using a shorter key space in Section No \ref{sec:s&p} and show that our proposed protocol provides sufficient security guarantees.

\begin{figure*}
    \centering
    \includegraphics[width=0.90\textwidth]{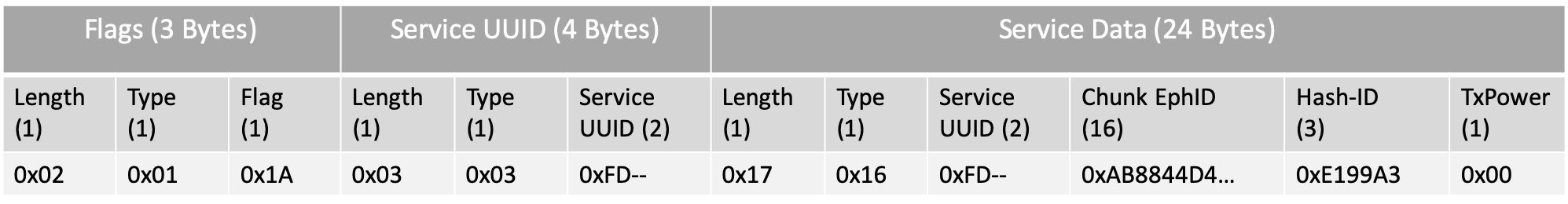}
    \caption{BLE advertisement message format.}
    \label{fig:BT}
\end{figure*}

\subsection{Advertising and receiving identifiers}
\label{subsec:adv}

%We have discussed in previous sections that we use contact identifiers to represent contacts between the devices and that devices are responsible for generating random identifiers that are advertised when these come in contact with each other. 
Once devices generate the $EphIDs$, the advertisement phase can commence. For this phase, instead of directly advertising the $EphID$, we use a $k$-out-of-$n$ secret sharing (Shamir's secret sharing)~\cite{sss} mechanism (explained in Section \ref{sec:k-out-n}). The device calculates $n$ secret part of the $EphID$ and broadcasts each share at the rate of 1 share advertisement per minute. A receiver can reconstruct the $EphID$ if it has successfully received any $k$ out of $n$ shares. For this work, we use the value of $k$ and $n$ as 15 and 30, based on the minimum duration of close contact defined as 15min by CDC.  %Reference were did we extract this numbers. (it is a footnote on intro)

%\begin{figure}
%    \centering
%    \includegraphics[width=0.48\textwidth]{figures/BloomFilter1.pdf}
%    \caption{Encoding encounter information into a Daily Bloom Filter.}
%    \label{fig:bloomfilter}
%    \vspace{-4mm}
%\end{figure}

There are multiple advantages of using this $k$-out-of-$n$ secret sharing. First, the devices need to be in contact for at least $k$ minutes to receive at least $k$ parts of the secret. Setting k = 15 min automatically takes care of the duration of close contact. %part of TC4TL (Too Close for Too Long). 
Any shorter contacts $< k$ minutes are not registered by the receiver. Additionally, we advertise part of the hash of the $EphID$ in each share. Figure \ref{fig:BT} shows the BLE advertisement message format with the 3-Byte hash of the $EphID$ included in the advertisement. This is to prevent the receiver from reconstructing the secret based on shares that are either less than $k$ or using shares advertised by different communicating parties. We note that even if the receiver uses an incorrect secret value for the encounter, this would never match with any other identifier at the contact tracing stage. In our version, a device will simply discard the shares, without attempting reconstruction, if it has not received at least $k$ advertisement of shares or if the hash values fail the integrity check of the reconstructed $EphID$. Thus, this mechanism increases the complexity for an adversary that is trying to capture the encounter identifiers for malicious use. A computationally bounded adversary, Eve, who is listening for BLE advertisements from Alice and Bob, has to first collect at least $k$ shares of the advertisements from both Alice and Bob. Then, she has to decrypt two random numbers from the reconstructed $EphIDs$ that would take significant time to compute due to discrete logarithm problems associated with the use of decisional Diffie-Hellman \cite{ddh}. 
%computationally bounded adversaries
%verifiable secret sharing
%%Can move this to attacks or privacy section

%If we consider Eve is able to compute the random numbers used by Alice and Bob, and consequently the encounter identifier, she can store this locally and in case she is diagnosed positive, both Alice and Bob will be flagged as close contact upon query. This, however, will not be an anomaly as Eve has indeed came in contact with Alice and Bob in the first place. The only way an attack can be launched is when Eve distributes this encounter ID (relating to Alice and Bob) to many malicious users (with modified app functionality) and they store it in their respective daily bloom filters.

\begin{figure}
    \centering
    \includegraphics[width=0.4\textwidth]{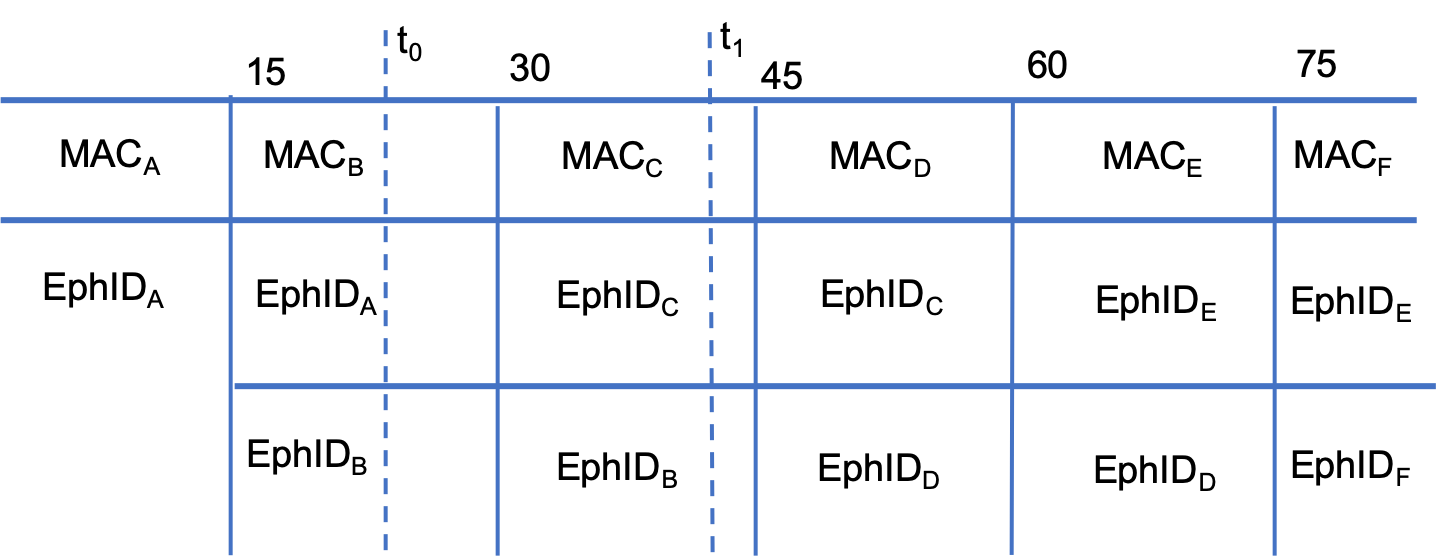}
    \caption{BLE advertisements with random MACs and $EphIDs$.}
    \label{fig:BT2}
    \vspace{-4mm}
\end{figure}
An additional issue associated with using \emph{k-out-of-n} secret sharing is the address carryover mechanism that is due to the rotation of the $EphID$ after each Epoch (30 minutes)\footnote{The Epoch is loosely aligned with the randomised MAC address periods that happen at approximately half of the Epoch duration.}. Suppose a receiver device B comes into contact with A when the 10th share of a particular $EphID$ is being broadcast by A at time $t_0$ and moves away when it has received the 10th share of the next $EphID$ generated by A at time $t_1$. Device B has thus maintained contact for 20 minutes, however, the logging mechanism would fail to register this contact as it has only received 10 chunks of each $EphID$.  

% can we serialise/order the id and send in circle (send the first 10 of last round again with next round)  --wanli

To address this issue, we use the simultaneous advertisement of multiple $EphIDs$ with overlapping intervals as proposed in \cite{contra}. A device always broadcast two $EphIDs$, rotating one identifier in such a way that the start of each identifier is staggered by 15 advertisement intervals. In Figure \ref{fig:BT2}, a device is broadcasting two overlapping $EphIDs$. A receiver device which it comes into contact with at time $t_0$ and leaves at time $t_1$ is able to register this contact as it has received enough shares of $EphID_B$ while in contact.

%Additionally, the two identifiers are advertised alternately for 30 sec. 
As a device is advertising shares of multiple identifiers, we also include the hash of the $EphID$, $Hash(EphID)$ truncated to the first 24 bits (3 Bytes) as the random identifier\footnote{We take the 128-bit randomly generated $EphID$, pass it through SHA-256 to get a 256-bit hash value, and then truncate it to retain the first 3 bytes.} that is used by the receiver to identify and combine different chunks belonging to the same $EphID$. Once a receiver has collected 15 shares of the same identifier, it reconstructs the identifier and verifies that it has received the correct identifier by computing the hash of the reconstructed ID and comparing the first three bytes with the hash included in the advertisements. Figure \ref{fig:BT} shows the message format of the BLE advertisement messages employed in our solution. We have used the ADV\_NON\_CONN\_IND message format for connectionless advertisements for the chunks of EphID. 

%add figure for BLE message exchange data format

%Shares advertised by a sender are differentiated by advertisements from other devices using the BT MAC addresses that we assume also get rotated in sync with one of the identifier.  

\subsection{Storing encounter information} 
\label{subsec:storage}

After sufficient shares of $EphIDs$ are exchanged with neighbouring devices, each pair of devices can compute a secret symmetric encounter identifier (referred to as $EncID$) only known to them. Each device inserts the encounter identifier in the local DBF. %(Figure \ref{fig:bloomfilter}). 
The computed encounter identifier is then deleted after it has been inserted into the DBF. We have used a Bloom filter to preserve the data privacy, reduce the storage requirement and improve the query efficiency, when compared with other normal data storage structures.

\textbf{Design of Bloom filter:}
The Bloom filter is a probabilistic structure that can result in false positives. The design of the filter involves various parameters such as the number of entries to be stored in the filter ($n$), the size of the filter in bits ($m$), the number of hashes ($k$) and the false positive rate ($p$).  
%%insert figures here%%
% \begin{figure*}
%     \centering
%     \begin{tabularx}{\linewidth}{XX}
%         \includegraphics[width=0.38\textwidth]{figures/fpr_dbf.pdf}
%         \caption{False positive rate for different number of encounters - DBF.}
%         \label{fig:FPR-DBF}
%         &
%         \includegraphics[width=0.38\textwidth]{figures/fpr_cbf.pdf}
%         \caption{False positive rate for different number of encounters - CBF. %\textcolor{red}{can we combine this two figure}
%         }
%         \label{fig:FPR-CBF}
%     \end{tabularx}
% \end{figure*}

% \vspace{-0.5inch}
\begin{figure}
    \centering
    \begin{tabularx}{\linewidth}{XX}
        \includegraphics[width=0.45\textwidth]{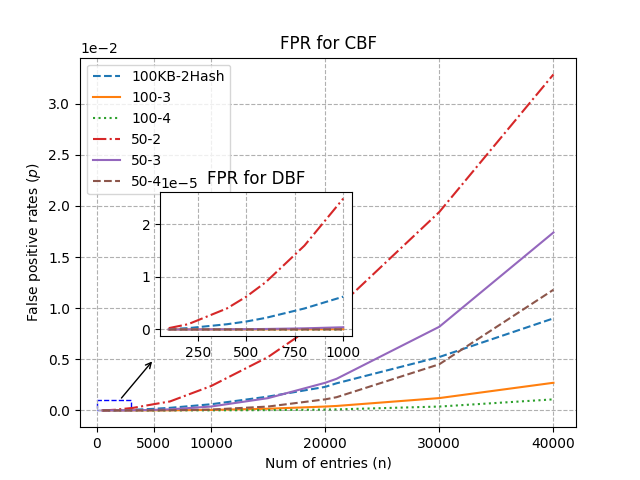}
        \vspace{-4mm}
        \caption{False positive rate vs number of encounters - DBF and CBF.}
        \label{fig:FPR}
    \end{tabularx}
\vspace{-4mm}    
\end{figure}
Figure \ref{fig:FPR} shows the false-positive rates for increasing values of encounters $n$ inserted in the DBF and CBF with different values of $m$ and $k$. Considering DIMY uses the secret-sharing mechanism (with at least 15 minutes to register a contact), we assume $n$ as 1,000 for DBF as a worst-case representing the maximum number of close contacts per day. As the CBF can hold a maximum of 21 DBFs, the worst-case $n$ for CBF is 21000. The FPR analysis shows that the worst FPR is given by $m$-$k$ combination of 50 KB-2 Hashes, and the best FPR by 100 KB-4 Hashes. As the hashing is performed at client devices, we take the size of the filter $m$ = 100KB with $k$ as 3 to reduce the computations and battery consumption. The DBF and CBF, in this setting, have FPR of 1 in 19 Million and 1 in 2303, respectively.   
%Following \cite{mitzenmacher2005probability}, %the optimal number of hashes to be used ($k$) comes out to be 2 using Equation \ref{eq:minifp}: 
%\begin{equation}\label{eq:minifp}
%k = \frac{m}{n} ln (2),
%\end{equation}
%The false positive rate $p$ comes out to be 0.0001 (1 in 6480) using Equation \ref{eq:fp}.
% \begin{equation}\label{eq:fp}
% p=(1-e^{-kn/m})^k
% \end{equation}
%repeated  to equation 1

%The DBF is utilised to create a Query Bloom Filter (QBF) to check for a possible exposure to positive cases, or to create a Contact Bloom Filter (CBF) representing contacts made by a positive case in the last 21 days.

\subsection {Uploading encounter identifiers to the blockchain}
\label{subsec:upload}
\vspace{2pt}
%i) Each device advertise a pseudo-anonymous random identifier that is saved by the receiving device. Once a user tests positive for COVID-19, they can either upload a list of their own used identifier for the last 2-3 weeks (approach taken by PACT-East Coast, DP-3T, EpiOne) or upload all captured identifiers for the last 2-3 weeks (used in CovidSafe and TraceTogether) ii) Represent the encounter with a unique identifier by combining a random identifier advertised by a neighbouring devices who have come in close contact. 
%push this Encounter identifier to their daily Bloom filter. 

%\subsubsection{Using Bloom Filter}
%\label{subsec:bloom}
%The device will store the received Encounters in the DBF as the local storage.  Each DBF\footnote{All Bloom filters (DBF, CBF and QBF) used in this design are of the same size} is designed with a 100KB size (i.e., length of 8000 bits) and the number of hash 2. Assuming the Encounter number needs to be stored is about 5,000, thus the estimate false positive rate is 0.0001 (1 in 6,480 chance). . Every day's Encounter will be stored in one Bloom filter, but if the encounter is less than \textcolor{red}{xx}, that day's Bloom filter will not be stored to avoid the \textcolor{red}{xx attack}.

% \begin{figure}
%     \centering
%     \includegraphics[width=0.48\textwidth]{figures/BloomFilter1.pdf}
%     \caption{Encoding encounter information into a Daily Bloom Filter.}
%     \label{fig:bloomfilter}
% \end{figure}

Once a user is diagnosed with COVID-19, they can get an authorisation code from the health authorities to upload their locally stored contact data to the back-end blockchain. Figure \ref{fig:flow} shows the information exchange (CBF and QBF to the backend and response from the backend) using a secure channel. The device combines their DBF covering the last 21 days into a single CBF of size 100KB (equal in size to the DBF). %Note that CBF has the same size (100KB) as DBF. 
The set union function is utilised as the combination process for the DBFs to construct a CBF. For example, all `1’-bit existing information in the DBFs are accumulated into one CBF by performing a bit-wise OR merging \cite{Papapetrou}. This merged CBF is theoretically equivalent to performing $\sum_{i=1}^{20} DBF_i \cup DBF_{i+1}$ and its false probability is similar to using a standard bloom filter. The CBF is then sent to the backend blockchain for logging as a transaction. The system supports querying by uploading of the QBF (encoded with the DBF from the last 21 days). The user's device uploads this query to the blockchain to check whether someone in close contact has tested positive.
%There is a two phase procedure for the Bloom filter encoding being used in the entire system. At the first phase, Bloom filter is mainly used to record the information and upload to the \textcolor{red}{blockchain}. At the second phase, user can actively check the exposure risk by querying the \textcolor{red}{blockchain} with their holding Bloom filters. 
%Once the user is tested positive (by the HO), \textcolor{red}{he/she can upload the accumulated CBF voluntarily}. It contains all encounter's information of the last 21 days. 

%The size of the CBF and DBF as 100KB has been chosen based on an analysis of the false positive rate (FPR) with the number of hashes fixed at 3.
%We take the worst-case scenario of a user recording 1000 close contacts each day for a continuous period of 21 days, resulting in the storage of 21 DBFs with 21,000 encounters in a single CBF. 
DBF, CBF, and QBF are of the same size of 100KB, serving three distinct purposes: First, it reduces the amount of data transferred to the backend, i.e., instead of 21 100KB sized DBFs, we only use one 100KB CBF or QBF. Second, this aggregation of multiple DBFs into a single CBF and QBF hides the details about the day of encounter to attenuate the privacy threat at the backend. Third, equal-sized CBF and QBF are employed to support efficient Bloom filter matching through set intersection operation at the backend (explained further in Section \ref{subsec:verify}).

%Therefore, the CBF equals to all data entries from the DBF are aggregated into the CBF without potential privacy information leakage threat, for example, the encounter date for a specific record, which help to attenuate the privacy threat on the back-end side.
%Figure~\ref{fig:FPR} shows that the CBF provides an FPR of 0.00261 (1 in 382) for 21000 encounters, although in practice the total number of encounters encoded in the CBF is expected to be much less, resulting in a lower FPR.
%\textcolor{red}{5000 comes from 15 mins and how many time sleep, how many people meet, multiple people}
%In the worst case,
%even if we consider that the user may meet and record the contact with double size 210,000, the number of hash $k$=2 can still achieves a false positive rate of 0.2:
%\begin{equation}\label{eq:minifp}
%k = \frac{l}{g} ln (2),
%\end{equation}
%leading to a false positive rate of 
%\begin{equation}
%f=\left(\frac{1}{2^{\ln %(2)}}\right)^{l / g},
%\end{equation}
%which is about 0.2. 
%Generally, the CBF would perform much better in terms of false positive rates as the number of encounters would be much less than the worst case assumed.  %fall into this high positive scenario (which the false positive rate should be $\le$ 0.0001). 
%However, any false positive reported by the system will suggest the individual to get cautious.
%\textcolor{red}{for CBL, even it is only one person, we record and accumulate but not allowed to do the check.}

%The blockchain performs the normal set intersection check which illustrates the overlap of two Bloom filters as an encounter information report.
\textbf{Design of Blockchain:}
%\label{subsec:BC}
We use Hyperledger Fabric~\cite{androulaki2018hyperledger} for the blockchain's implementation, as it provides a modular permissioned blockchain platform, which allows flexibility in modelling the Bloom filter on transactions. The Hyperledger Fabric network is designed to be maintained by a consortium of Health Authorities (HAs) which comprises of stakeholders in the healthcare sector, e.g., relevant government agencies and hospitals. Each HA maintains a set of peer nodes to host the ledgers, execute smart contracts, and maintain a set of orderer nodes for consensus protocol. HAs and their corresponding peers are identifiable by cryptographic primitives that comply with the X.509 standard for public-key certificates. 

The HAs interact with the blockchain through multiple smart contracts. We have designed a smart contract that is capable of performing the following functionalities:
\begin{itemize}
    \item \textit{Issuing access tokens}: Only users who test positive are allowed to upload their CBF to the blockchain. Each user who has tested positive is given a temporary token by the HA that authorises them to access the back-end. The HA transacts with the blockchain to issue the temporary access token by providing corresponding HA credentials to the smart contract. Upon successful credential validation, the smart contract records the token to the blockchain. Note that, this temporary token is only valid for 24 hours.
    \item \textit{Processing CBF}: The smart contract validates the temporary access token provided by the user who uploads their CBF. Upon successful validation, the smart contract records the CBF permanently to the blockchain and updates the ledger's state.
    \item \textit{Processing QBF}: The smart contract handles queries from users concerning contacts with positive cases by checking the user's QBF against stored CBF in the ledger. Then, the smart contract returns the matching result, which will either be true or false.
\end{itemize}
\noindent

\begin{figure}
    \centering
    \includegraphics[width=0.48\textwidth]{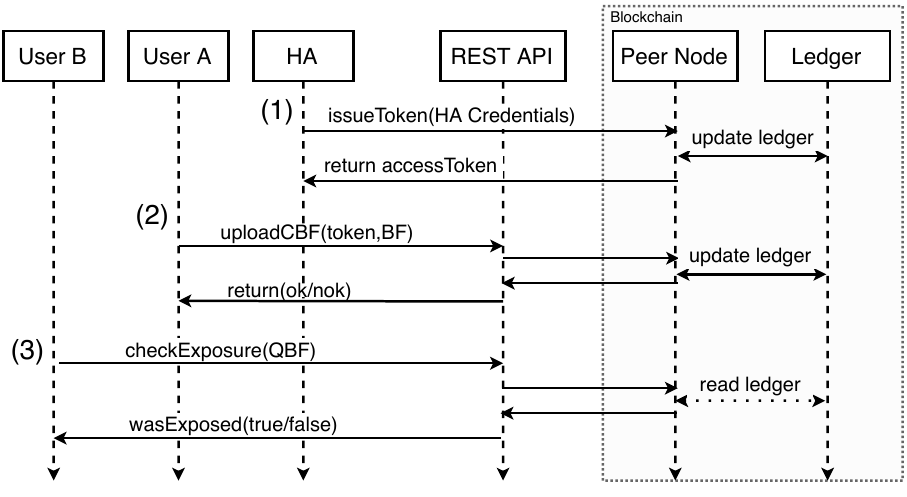}
    \caption{Uploading to the blockchain.}
    \label{fig:upload-to-blockchain}
\end{figure}

%The blockchain back-end to support the solution relies in the Hyperledger implementation. Hyperledger Fabric 
% -Chaincode (Smart contracts) are used to: 
% i) validate the token sent in the transaction is valid, i.e., it was issues by a health authority and did not expire.
% ii) create a new block composed of multiple bloomfilters (CBL)
% iii) receive a query transaction (QBL), that contains a bloomfilter, and compare with the existing 

% API for communication with users' devices
CBFs stored at the blockchain are managed and queried by the Hyperledger. Given the on-chain data storage capacity of a single transaction is 4MB~\cite{hyperledgerTransactionSize}, the Hyperledger can add a minimum of 1 or a maximum of 40 CBFs (40x100KB = 4MB) in a single transaction. %The Bloom filter applied to store the transactions makes possible to the solution requiring a small storage space (in comparison to traditional full encounter ID exchanges storage).

To ensure the user's anonymity, interaction with the blockchain is provided only through REST APIs. %Therefore, authentication is not required, and the user can send their queries directly. Note that, 
To upload the CBF, users need to include their temporary access token with the query. The query mechanism does not require any access authentication. The REST APIs are provided by HAs, which imply that multiple identical APIs are available.

Figure~\ref{fig:upload-to-blockchain} shows the process of uploading relevant information to the blockchain, along with the main actors involved. In general, the interaction involves three sub-processes. Step 1) The HA issues a temporary access token by providing corresponding HA credentials to the blockchain in a transaction via a peer node. The peer node validates the credentials and logs the transaction in the blockchain to mark the issuing of a token. The HA receives the temporary access token and transmits the token to the appropriate user (User A in this case). Step 2) After receiving the token, User A may upload its CBF using the REST API with the appropriate access token. The REST API then forwards the request and responds with the transaction status. Step 3) In an event in which User B wishes to check for potential contact with positive cases, User B can interact with the REST API with a \texttt{checkExposure(QBF)} message, which includes its QBF. The REST API forwards the query to the blockchain for checking if the QBF contains potential contacts with positive cases. The REST API messages \texttt{wasExposed} to User B to indicate that User B was in contact with an identified positive case.

\subsection{Contact verification process}
\label{subsec:verify}
The contact verification process is performed through a smart contract at the blockchain. Each day\footnote{For scalability, the query is performed in a 24-hour cycle from the time of app installation.} the app combines all the DBFs into a single QBF. The user also appends the date of the oldest Bloom filter, $T_{old}$, used in the query. The blockchain takes this query and runs a search through the blocks, trying to match any entry in the QBF with existing CBF transaction entries. Note that the search is restricted to only transactions following the $T_{old}$ date\footnote{$T_{old}$ date can be a maximum of 21 days old, thus any CBF stored at the blockchain that is older than 21 days is not matched. This automatically takes care of CBFs pertaining to COVID-19 cases that are no longer infectious.}. This search equates to finding the  intersection of the two equal-sized filters CBF and QBF constructed with equal number of hashes. This is done by performing a bitwise-AND operation on CBF and QBF to approximate their set intersection. Let $t$ denote the number of bits set in the intersection set, FPR for the intersection set is $= (t/m)^k$. Since $t$ is always less than or at most equal to the set bits in any of CBF and QBF, the FPR for intersection set is $\leq$ FPR for CBF, and is $\leq$ FPR for QBF \cite{Papapetrou}. The blockchain returns the appropriate response matched or not matched based on the number of set bits in the intersection set.

\begin{table*}[ht]
\vspace{-4mm}
  \footnotesize
  \centering
  \caption[Compare]{Comparison of DIMY with other protocols}
  \vspace{-4mm}
  \label{table:Compare}
    \begin{tabular}{ |c|c|c|c|c|}
      \hline
      \rowcolor{Gray}
      Salient & DIMY &Centralised &Decentralised  & Hybrid\\
      \rowcolor{Gray}
      Features &  & (BlueTrace)& (PACT-East, DP-3T)&(Desire)\\
      %&  & &&\\
      \hline
      ID generation& Client devices & Server & Client devices & Client devices\\
      %generation &&& &\\
      \hline
      Storage on & Encounter encoded in & Received IDs from &Received IDs (chirps)  & EphIDs and\\
       devices & Bloom filters &close contacts &from close contacts  & two PETs tables\\
      \hline
      Storage on & Encounter encoded in &Mapping of IDs, Complete list &Hourly seeds and  & PETs for\\
      server/back-end&Bloom filters for positive cases & of contact IDs for positive cases  &time for positive cases  &positive cases\\
      \hline
      Processing on & ID generation, Diffie-Hellman& Minimal processing&  Hourly seed and chirp& ID generation\\
      devices &key generation, $k$-out-of-$n$ secret& &generation, Chirp matching  & Diffie-Hellman\\
      &sharing, Bloom filter encoding&  & and risk analysis  & exchange\\
      \hline
      Processing on& Blockchain matching&  Risk analysis and&Minimal processing  &Risk analysis \\
      server/back-end&for at-risk users& ID matching&  &PETs matching \\
      \hline
      Data upload & Bloom filter for positive cases &All contact IDs captured &Seeds, timing information & PETs table\\
      & Query Bloom filter for other users &for a positive case &for a positive case & for positive case\\
      \hline
      Data download & Result (yes/no) &  Periodic download of&Seeds,timing information& Result of\\
      &from blockchain& new IDs&for all positive cases  & risk analysis\\
      \hline
      Risk Analysis& Performed on &Performed on &Performed on & Performed on\\
      \& notification& Blockchain & server& devices& server\\
      \hline
    \end{tabular}
    \vspace{-4mm}
  \end{table*}

\section{Comparison}
\label{sec:Comp}
We introduced three architectures commonly used for designing digital contact tracing apps in Section \ref{sec:rw}, and discussed the design of our proposed solution in the previous section. In this section, we compare the salient features of our proposed solution with representative apps from the three architectures. 

Table \ref{table:Compare} highlights the salient features and their equivalent in selected apps. Our proposed solution, DIMY, generates a temporal ID on the client's device in line with other decentralised and hybrid apps. %This is done to safeguard against disclosure of linking information to the back-end/server. 
DIMY is also optimised for storage, both on the client's device as well as the back-end. The design involves storing contact information in fixed-size DBFs. The back-end blockchain only stores a single Bloom filter (CBL, size 100KB) per positive case that has encoded information on the DBFs for the last 21 days. This reduces the storage requirement at the back-end/server considerably when compared with other apps listed in Table \ref{table:Compare}.

Another salient design question for digital contact tracing apps is where to perform the risk analysis and notification. Apps based on centralised and hybrid architectures perform this step at the centralised server, while apps based on a decentralised architecture perform this locally, on the device. DIMY performs the matching of contact information on the back-end blockchain. However, the blockchain is not able to infer any extra information as the matching is performed on contact information encoded in Bloom filters.

On the other hand, our proposed solution is device-centric in the sense that it performs most of the privacy-preserving operations on the device. This includes EphID generation, computing $k$-out-of-$n$ shares and broadcasting these shares using BLE messages, and encoding received contact information on DBL after enough shares have been received to construct a shared secret. In comparison, apps such as TraceTogether and CovidSafe (AU) only involve the advertisement of IDs received from the centralised server. Desire also uses the Diffie-Hellman key exchange and the generation of local IDs on the devices. In contrast, Desire uploads the shared secrets collected by a user who has been diagnosed with COVID-19 to a server for matching, to be performed at a later stage. 
%Could we quantify the amount of data upload?
DIMY requires uploading the least amount of data when compared with other apps. A single 100KB sized CBL is uploaded from a COVID positive  client to the blockchain. This is in contrast with uploading all contact IDs, required on apps that follow the centralised architecture, and uploading multiple seeds or the PETs table on apps that use decentralised and the hybrid architectures. DIMY also requires client devices to upload QBL, a Bloom filter for matching transactions stored on the blockchain. DIMY client devices only download the results of the risk analysis in the form of a binary notification similar to centralised and hybrid apps. Apps based on the decentralised architecture involves the downloading of either seeds/chirps from the server in order for matching to be performed on the devices.    
%Compare with centralised, decentralised, hybrid
%Tabular feature comparison
\section{Security and Privacy Analysis}\label{sec:s&p}
This section is dedicated to an analysis of security and privacy guarantees provided by the DIMY design of protocol.
\subsection{Threat Model}
\label{subsec:threat}
In this section, we describe the adversaries considered in the design of the DIMY protocol and the risks that they pose. We categorise the adversaries into three groups, users, back-end developers/administrators and the government. i) \emph{Users} have access to in-app information as well as passive information captured through eavesdropping. App users are also assumed to have access to the open-source app code. Furthermore, we assume users can only have access to data stored on other smartphones through theft or coercion. ii) \emph{Backend administrators/developers} have access to all data received and stored at the backend server. %Additionally, they have access to other public information related to contact tracing. 
iii) \emph{The government} can access any information stored on individual smartphones or the backend server through subpoenas to investigate a group or individual user of the app. 
\subsection{Security and Privacy Analysis}
%%Risk of all data on the device being compromised%%
%%Faking risk by uploading malicious IDs of other users%%

Requirements: 

\begin{enumerate}
	
	\item Completeness: If a person  receives an alert message ``match", then he/she was definately in proximity to a COVID positive patient. 
	\item Soundness: If a person was not in the proximity of a COVID positive patient, then there is only a negligible probability that he/she will receive an alert. 
	\item Type I Privacy: No information is revealed about a patient who has tested positive.
	\item  Type II Privacy: Even if the data present on a device is compromised, it does not leak information about individuals who were in close proximity of the device.             
\end{enumerate}

%\begin{theorem}
	If the key exchange and secret sharing mechanisms are secure, and the Bloom filter and the blockchain implementation are correct, then the protocol achieves completeness, soundness, Type I and Type II privacy. 
%\end{theorem}

%\begin{proof}

\textbf{Completeness:}

	A user who wants to check whether he/she was in close proximity with a COVID positive individual sends their QBF to the back-end. This is matched against the CBF on the blockchain. Since the check is performed by a smart contract run by peer nodes, it returns a match only if a match exists. This is ensured by the set membership propoerties given by the Bloom filter and the assumption that the blockchain implementation is correct. We note that the QBF is sent through a secure channel. 
	
\textbf{Soundness:}

	An attacker who was not in close proximity with a COVID positive patient will receive a match with negligible probability (as computed by Equation \ref{eq:fp}). This is also ensured by the property of the Bloom filter and the assumption that the blockchain implementation is correct. 
	The blockchain stores only CBFs, Bloom filters that encode the encounter identifiers of COVID positive patients.
	An attacker, who was not in proximity with a COVID positive patient, cannot construct an encounter identifier for a QBF that will match with the CBF. This is because of the properties of decisional Diffie-Hellman and the hash functions used for encoding encounters in the CBF.         
	
	%An attacker cannot receive a "match" if it was not in proximity of a covid positive patient for the following reason.  In order to match,  should have atleast one encounter identifier $g^{X_{At}Y_{Bt}}$ matched with a CBF. Encounter identifier  cannot be constructed from  CBF. To construct a valid encounter identifier, it should possess at least $k$ shares of $g^{X_{At}}$ or $g^{Y_{Bt}}$. Even if it is able to construct  $g^{X_{At}}$ or $g^{Y_{Bt}}$, it cannot construct $X_{At}$ or $Y_{Bt}$, because of the hardness of the discrete logarithm. 

	\textbf{Type I privacy:} 
	
	The blockchain stores only CBF. A user who uploads a QBF to the blockchain and receives a match knows that he/she has been in contact with at least one COVID positive patient but cannot say which one, due to the design of the Bloom filters. Here we ignore the case in which a person has been in contact with only one person during the last 21 days or uploads only one entry QBF and receives a "match". In such a case, the identity of the COVID positive patient is known.  
	
	\textbf{Type II privacy:} 
	
	When a  device is compromised, the DBF is revealed. Since the \emph{EncID} and  \emph{EphID}s are not known, and the secrets  corresponding to the \emph{EphID}s of the device $X_{At}$ are not known, the attacker cannot know the identity of users in close proximity with the device. 
	
%\end{proof}

% \begin{table*}[ht]
%   \footnotesize
%   \centering
%   \caption[Attacks]{Possible attacks on digital contact tracing}
%   \label{table:Attacks}
%     \begin{tabular}{ |c|c|c|c|c|}
%       \hline
%       Attacks & DIMY & Centralised & Decentralised & Hybrid\\
%       &  & Architecture & Architecture& Architecture\\
%       &  & (BlueTrace) & (PACT-East, DP-3T)& (Desire)\\
%       \hline
%       Replay Attack & $\times$ & \Checkmark & \Checkmark & $\times$\\
%       \hline
%       Relay Attack & \Checkmark & \Checkmark &\Checkmark  & \Checkmark\\
%       \hline
%       Device Tracking & \Checkmark &  \Checkmark&\Checkmark  & \Checkmark\\
%       \hline
%       Carryover Attack& \Checkmark &  \Checkmark&\Checkmark  &$\times$ \\
%       \hline
%       Location Confirmation& $\times$  & \Checkmark &$\times$  & $\times$\\
%       \hline
%       Enumeration Attack & $\times$ &  $\times$&\Checkmark  & $\times$\\
%       \hline
%       Denial of Service& \Checkmark &  \Checkmark&\Checkmark  & \Checkmark\\
%       \hline
%       Deanonymization/Linkage & $\times$ & \Checkmark &  \Checkmark&$\times$ \\
%       \hline
%       Social Graph & $\times$ & \Checkmark & \Checkmark & \Checkmark\\
%       \hline
%     \end{tabular}
%   \end{table*}

\begin{table}[ht]
  \footnotesize
  \centering
  \caption[Attacks]{Possible attacks on digital contact tracing (C=Centralised, D=Decentralised, H=Hybrid)}
  \vspace{-4mm}
  \label{table:Attacks}
\begin{tabular}{|c|c|c|c|c|}
\hline
Attacks                                                              & DIMY & \begin{tabular}[c]{@{}c@{}}C\\ (BlueTrace)\end{tabular} & \begin{tabular}[c]{@{}c@{}}D\\ (PACT-East,\\ DP-3T)\end{tabular} & \begin{tabular}[c]{@{}c@{}}H\\ (Desire)\end{tabular} \\ \hline
Replay                                                               & $\times$    & \Checkmark                                                       & \Checkmark                                                               & $\times$                                                    \\ \hline
Relay                                                                & \Checkmark    & \Checkmark                                                       & \Checkmark                                                               & \Checkmark                                                    \\ \hline
\begin{tabular}[c]{@{}c@{}}Device \\ Tracking\end{tabular}           & \Checkmark    & \Checkmark                                                       & \Checkmark                                                               & \Checkmark                                                    \\ \hline
Carryover                                                            & \Checkmark    & \Checkmark                                                       & \Checkmark                                                               & $\times$                                                    \\ \hline
\begin{tabular}[c]{@{}c@{}}Location\\ Confirmation\end{tabular}      & $\times$    & \Checkmark                                                       & \Checkmark                                                               & $\times$                                                    \\ \hline
Enumeration                                                          & $\times$    & \Checkmark                                                       & $\times$                                                               & $\times$                                                    \\ \hline
\begin{tabular}[c]{@{}c@{}}Denial of\\ Service\end{tabular}          & \Checkmark    & \Checkmark                                                       & \Checkmark                                                               & \Checkmark                                                    \\ \hline
\begin{tabular}[c]{@{}c@{}} Linkage\end{tabular} & $\times$    & \Checkmark                                                       & \Checkmark                                                               & $\times$                                                    \\ \hline
Social Graph                                                         & $\times$    & \Checkmark                                                       & \Checkmark                                                               & \Checkmark                                                    \\ \hline
\end{tabular}
\vspace{-4mm}
\end{table}

\subsection{Privacy Protection}
In this section, we discuss the DIMY's privacy protection properties. In our solution, we adopted blockchain as the back-end service. %, following the commonly adopted `honest-but-curious' server assumption for the threat model. 
As sensitive data covering encounter information is first encoded in Bloom filters by clients before uploading to the chain, there is implicit privacy protection against possible back-end breaches (discussed earlier in the threat model). The one-way hash mapping involved in the Bloom filter also significantly reduces the back-end's ability to construct social graphs based on the diagnosed user's contacts. The only information the back-end can infer is an estimate of the number of contacts encoded in the CBL uploaded by a positive case,  without identifying who these contacts are. On the other hand, users can query the blockchain for matching using QBLs. The result of the query is a simple binary decision that does not reveal which of the encounters in QBL have matched. As users do not have direct access to the CBL stored in the blockchain, they are unable to extract any sensitive information.

%All information stored on the block can be publicly accessed. All users can access the Blockchain for query purpose for example, however, only the user with locally stored information (QBL, datetime, etc) can interpret the query result. Malicious user who even got a matching query result cannot retain any useful information. 

 %Only the user who holds the same matching pattern will get informed about the possible risk information. 
 The devices, on the other hand, only broadcast shares of the cryptographically generated EphID via BLE advertisements in the contact phase.  These pseudo-identifiers cannot be matched back to the real identities of the users unless %generated by the cryptographic based method which will only work as the pseudo-identifier but nothing. As long as those pseudo-identifiers cannot be matched back to the real user identity/smartphone (this is detailed in the security analysis section~\ref{sec:s&p}), there won't be any extra information leakage.
 an adversary accumulates a significant amount of auxiliary information, such as location information obtained by hacking GPS, or eavesdropping on WiFi or other sensors. %, undoubtedly, the user's identity could be revealed and traced. However, the proposed protocol will not reveal extra information which could be used link the user's identity with diagnose report. Only if the smartphone is hacked and controlled, in which all CBL are noted down as long as its generated surroundings, the attacker can map back the captured information with the information on the Block. But in this case, there is no any privacy space existing.
All encounter information is deleted once it is encoded in the DBF, hence protecting data in case a device gets physically stolen or the user is forced to reveal app data under coercion. 
%%Wanli%%
%%Discuss function creep%%

Another potential privacy concern around contact tracing apps is known as the function creep \cite{DP3T}. Function creep refers to the evolution of the app to include functionalities other than the original ones, i.e., the app has the potential to be turned into an instrument of mass surveillance, violating human rights. Thus, it is necessary to analyse the privacy of the proposed app from a function creep perspective. 

In the proposed protocol, temporary IDs (i.e., EphIDs) generated by devices are first used to construct encounter IDs that are then encoded in the Bloom filter, which stops underlying linkages from being created between the temporary IDs and concrete IDs in the real world. %such as smart travel card or CCTV footage. 
This binary data encoded in a Bloom filter becomes semantically meaningless to any other user, and even the back-end cannot associate the reported data with an infected person or any specific individual. %Further, the proposed app only permits the user to query the back-end database stored on the chain instead of server sending e to the user. %This capability can disable the third party (e.g., law enforcement) to trace the movements of user and communities by assigning them distinguishable identifiers (via notification message for example) and recognising their tagged Bluetooth emission unless keep following the target physically. As a conclusion, it is nearly impossible to turn the proposed app into a potential surveillance tool.

Lastly, %corrupted participants are also playing as key factors in various contact tracing apps. For 
in the proposed protocol, the blockchain is adopted on the back-end, which provides transparency on the integrity and trustworthiness of the data stored on the chain. %can guarantee the `server' can never be corrupted. 
Thus, the server is unable to extract any extra information that could assist a compromised back-end to use the stored data for any other purpose. However, the DIMY protocol is also susceptible to privacy attacks launched by malicious users who may use a modified application to collect other contextual information regarding the contacts. %A malicious user implies that she may be able to modify the application (or use a tool) in order to collect the additional data not typically collected. 
Multiple malicious users may work together, combining their information, to collect a large number of recorded broadcasts with metadata on time and location, etc. %With the benefit of Bloom filter masking and non-notification server design, the adversary can only know $m$ users' \textit{real identity} and  \textit{diagnosed risk/exposures} (assuming there are $m$ corrupted users), the $m+1$ privacy information is still unknown. Thus, the adversary cannot infer other users' information related to those two factor. However, DIMY as a BLE-based contact tracing protocol also suffers from the  \textit{location privacy} issue. Even without any side-channel information, corrupted user can infer non-corrupted user's location history by accumulating the meta data while broadcasting, which depends on the amount of available background information the adversary may have about the movement pattern and the estimation of resources, e.g., Bluetooth receivers (corrupted users) needed to cover the assumed movement area.

%%TODO%%
%%Add one entry attacks%%
%add one record only in BF every time

%%Backend deploying own tracking devices%%
%still corrupted user with EphID recorded if hacked phone

%\subsection{Formal analysis using AVISPA}
\subsection{Resilience against attacks}
\label{subsec:resilience}
In this section, we will explore the resilience of our proposed design against common attacks launched against digital contact tracing apps.

\subsubsection{Replay attacks}
In this type of attack, the goal of an adversary is to inject malicious contact information entries such that these entries result in false positives\footnote{A false positive occurs when the digital app indicates a close contact with a positive case, even though that contact has not occurred.} during the contact verification stage. An adversary can capture the advertised messages by a user's device and replays these at another location later on. 
Our proposed solution provides a safeguard against such attacks by using the secret sharing scheme. An adversary must capture at least $k$ shares of a message, before taking these shares to another location for rebroadcasting. This may result in several recipients using these shares to form contact information in their logs. However, in order to be counted as false positives, the originator of the messages must also have matching entries in their logs. The only way this attack would be possible is if the adversary moves back and forth between two different locations, collecting shares and rebroadcasting these to ensure the existence of symmetric contact information. 

The design of our proposed solution renders it impossible, as the adversary has to remain at a particular location for at least $k$ minutes before moving to another location to advertise these collected shares. Assume that the adversary can collect $k$ shares advertised by Alice at location 'A', and rebroadcast it at location 'B'. Once the adversary returns to location 'A' with the shares collected from location 'B', the advertised $EphID$ for user Alice has changed (it has a rotation time of 30 minutes), so it would not result in the storage of symmetric contact information.  

\subsubsection{Relay attacks}
An adversary's objective during a relay attack is the same as it is in a replay attack. An adversary can capture a user's advertised shares and immediately relay the captured message at the same location, extending the range of the message. The adversary thus acts as a relay, transmitting shares that it manages to capture.

Our proposed solution is susceptible to relay attacks that are inherent to all schemes using BLE messages. It is possible to rebroadcast shares such that two users, Alice and Bob, have symmetrical contact information even though they were not in direct contact with each other. We point out that both Alice and Bob have to be in direct communication range of the adversary for $k$ minutes to obtain symmetric contact information. As a consequence, if either Alice or Bob tests positive, the other user would be informed of a `false positive' close contact.

\subsubsection{Device tracking}
The adversary’s goal in this type of attack is to exploit the fact that most digital tracing apps use BLE and BLE information broadcasts can be used to track a particular device. A passive listening device can listen for BLE advertisements/connections and transfer these to a central tracking server. The server can diffuse information from multiple tracking devices to estimate the position and movement pattern of the device being tracked.
It is thus trivial to track a device that is advertising BLE messages while there is an identifier that can be associated with that device. In regular communications, the Bluetooth MAC address is randomised for a short period to limit this tracking. In our proposed solution, we use chunks of $EphIDs$ that are different from each other and use the $EphID$ hash to link all these shares together. An adversary can use the $EphID$ hash in combination with the randomised MAC address to perform limited tracking.

\subsubsection{Location confirmation}
The adversary’s goal is to discover the presence of a user at a known location. This is accomplished by linking contextual information, such as the mobile phone model used in BLE advertisements in apps that are based on centralised contact tracing architectures. This type of attack is not possible in our proposed protocol, due to the use of ephemeral identifiers and the suppression of other information that links the device with a particular user.    
\subsubsection{Enumeration Attack}
This attack aims to estimate the number of users who have uploaded their contact tracing data after testing positive with COVID-19.
In our proposed protocol, the encounter data is first encoded in Bloom filters before being stored on the blockchain. A user is allowed to query the blockchain for matching any encounter record, without revealing the records stored on the blockchain. A malicious user  is thus unable to launch an enumeration attack. Note that as the HAs authorise all uploads of CBFs to the blockchain, and there are multiple HAs that exist in the system, they can collude with each other to arrive at the total number of COVID-19 cases that have uploaded CBFs to the blockchain.      
\subsubsection{Denial of Service}
In this type of attack, an adversary generates fake advertisements to consume the storage and battery resources of other devices. Digital contact tracing apps are prone to this attack irrespective of their underlying architecture. In our proposed solution, an adversary can force other devices to store fake encounter information by advertising multiple $EphIDs$ instead of using only two identifiers.
%, if it can also associate multiple random BT MAC addresses with these advertisements. 
\subsubsection{Deanonymisation/Linkage}
An adversary aims to deanonymise a user based on the information it can collect either through the system or by using a side-channel. This attack can be launched to deanonymise close contacts or to identify users who have tested positive. This type of attack is not possible in our proposed solution as information regarding close contacts and positive cases is not directly shared with other users. The query mechanism through the blockchain does not reveal details of an encounter with a positive case; rather, it simply informs the affected user that they are at risk. 
%\subsubsection{Abuse of app}

\subsubsection{Carryover attack}
An address carryover attack is possible when the changeover time of a randomised Bluetooth MAC address and the temporary identifier are not synchronised. A listener can thus easily link the multiple Bluetooth MAC addresses advertised within the same identifier's life time. Our proposed solution relies on the simultaneous advertising of two identifiers to enable the correct contact information to be captured (discussed in Section \ref{subsec:adv}). This mechanism may result in a carryover attack for tracking purposes, whereby an adversary can associate multiple advertised identifiers with the BT MAC addresses used by a device. %Figure (\ref{fig:carry}) illustrates the carry over attack. 
An adversary can associate the $EphID$ hash that is being advertised along with random MAC and chunks of the raw $EphID$ to track a user's device, as long as that device is within the communication range.    
\subsubsection{Social Graph Analysis}
Social graph construction enables the identification of a person's close contacts. This is an imperative part of manual contact tracing in which a health official conducts interview with a positive case to identify their at-risk close contacts. Digital contact tracing mechanism stores the contact information locally on the user's device. This information is utilised to identify close contacts once a user tests positive. %Construction of social graph is trivial in the centralised architecture where the server has access to all user IDs and their close contacts. In the decentralised architecture, the IDs are generated by the user devices hence the social graph identification is hard without side-channel information. 
In our proposed solution, we have employed two mechanisms that prevent the construction of social graphs. First, we have made use of ephemeral ID generation on the devices as opposed to ID generation by the server. This means that the back-end blockchain cannot link an $EphID$ with a user. Second, we have employed Bloom filters to hide the contact information of positive cases from the distributed blockchain. The blockchain is thus unable to construct a social graph when a user either uploads their contact Bloom or queries the blockchain using the QBF.

Table \ref{table:Attacks} summarises this section with details of the attacks that could be launched against various architectures, including against our proposed design.

\section{Performance evaluation}
\label{sec:evaluation}
\vspace{2pt}
In this section, we present a quantitative evaluation of our proposed backend solution based on  blockchain implementation, in terms of throughput, latency and resource consumption. We note that for these experiments we generated synthetic data on the device level and supplied this to the blockchain. Our implementation of the DIMY app and its GUI is part of our future work.

% \begin{figure}[!htb]
% \minipage{0.32\textwidth}
%   \includegraphics[width=\linewidth]{figures/task_compare.stdev.pdf}
%   \caption{A really Awesome Image}\label{fig:awesome_image1}
% \endminipage\hfill
% \minipage{0.32\textwidth}
%   \includegraphics[width=\linewidth]{figures/cpu_mem.pdf}
%   \caption{A really Awesome Image}\label{fig:awesome_image2}
% \endminipage\hfill
% \minipage{0.34\textwidth}%
%   \includegraphics[width=\linewidth]{figures/varying_qbf.pdf}
%   \caption{A really Awesome Image}\label{fig:awesome_image3}
% \endminipage
% \end{figure}

\begin{figure*}
    \centering
    \begin{tabularx}{\linewidth}{XXX}
        \includegraphics[width=0.32\textwidth]{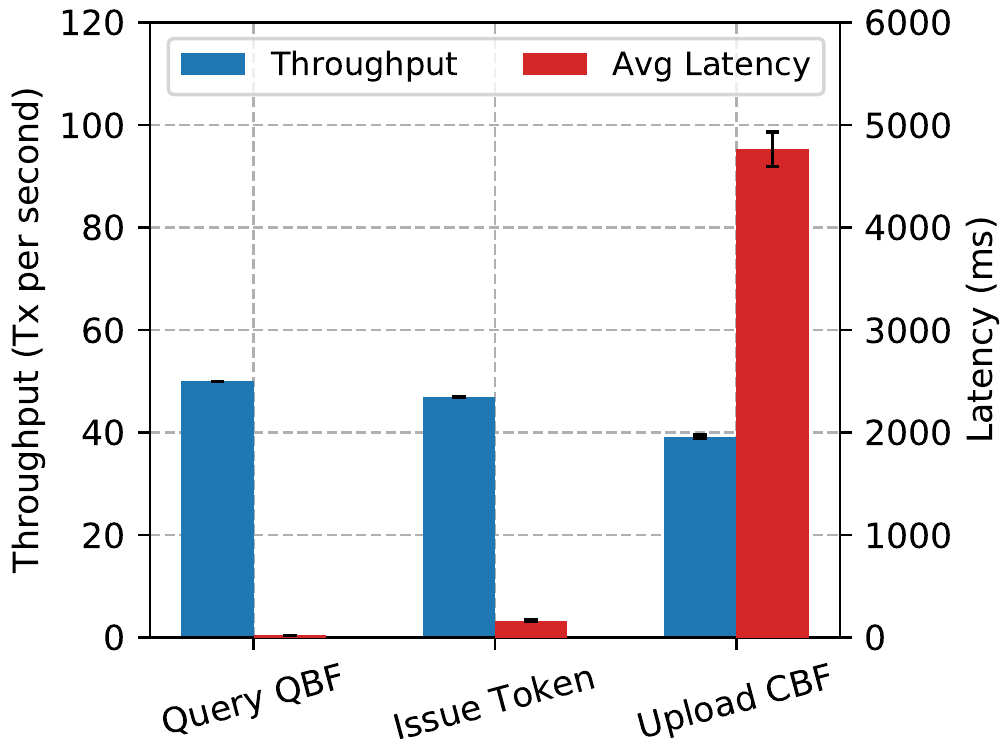}
        \caption{Comparison of the average latency and the throughput for blockchain operations in caliper, using a load of 50 transactions per second.}
        \label{fig:task-compare}
        &
        \includegraphics[width=0.32\textwidth]{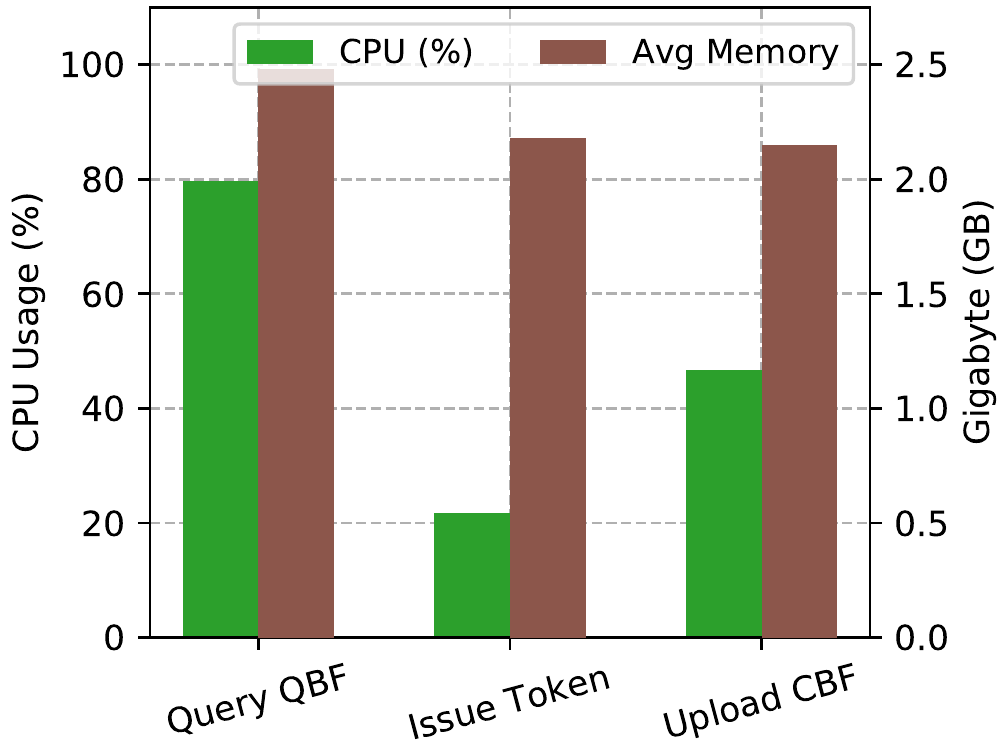}
        \caption{Comparison of CPU and memory consumption for the execution of querying QBF, issuing an access token and uploading CBF.}
        \label{fig:cpu-mem}
        &
        \includegraphics[width=0.32\textwidth]{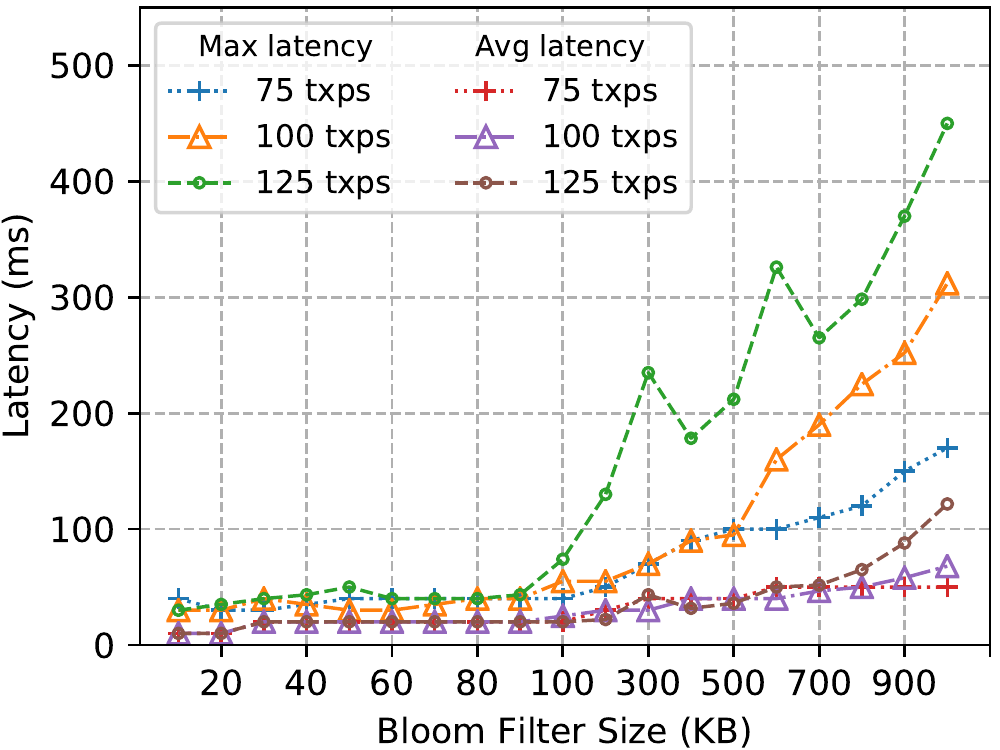}
        \caption{The maximum and average latency of querying QBF with differently sized Bloom filters.}
        \label{fig:varying-qbf}
    \end{tabularx}
\end{figure*}
% \vspace{-0.5inch}

\begin{figure}
\centering
\includegraphics[width=0.41\textwidth]{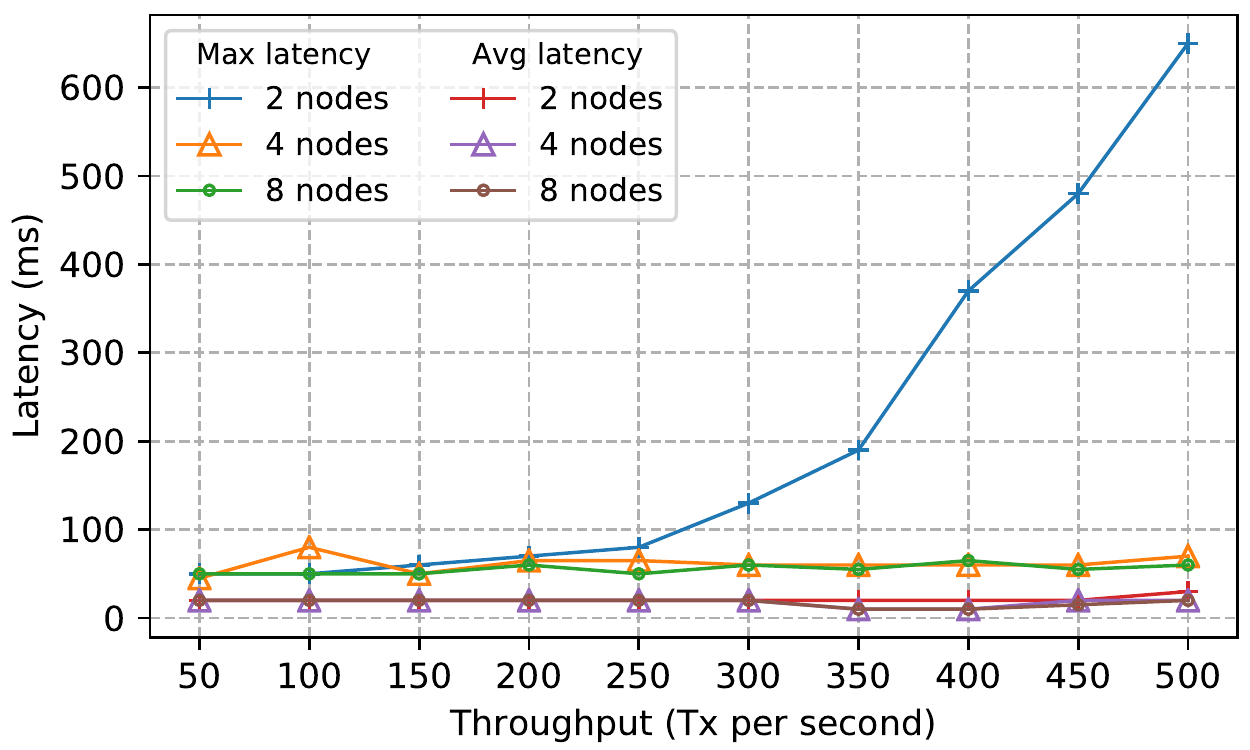}
\caption{Throughput and latency of querying QBF with the size of 100KB in different transaction send rates and number of Hyperledger peer nodes.}
\label{fig:throuhgput}
\vspace{-4mm}
\end{figure}

\subsection{Implementation details}
\label{subsec:poc}
\vspace{1pt}
    We implemented a proof-of-concept of our proposed framework using Hyperledger Fabric v2.0, as it allows flexibility in modelling Bloom filters in a permissioned blockchain environment. We opted to use a permissioned blockchain to control the app user's access by regulatory organisations, such as the health authority.
    %Additionally, it enable to model a network comprised of two regulatory organisations, namely a government's agency and a health authority. 
    We consider the standard configuration of a solo orderer node with one communication channel as the consensus mechanism. We ran our experiments on a single machine with 12 cores of CPU and 64GB of memory, running Ubuntu Linux 18.04 LTS. We implemented the core functions of DIMY transactions as chaincodes written in the Go programming language. We selected a native Go implementation of the Bloom Filter v2.0.3\footnote{https://github.com/willf/bloom} and a non-cryptographic murmur hashing function for Go %\footnote{\url{https://github.com/willf/bloom}} 
     to implement the Bloom filter functionality in the Hyperledger Fabric. We benchmarked our proof-of-concept implementation using Caliper v0.3.2\footnote{https://hyperledger.github.io/caliper/}, an official tool from the Hyperledger foundation that allows blockchain designers to measure the performance of the implementation of a specific blockchain. We measured the performance per second and repeated the measurements for 30 seconds. We noted the performance of our implementation in terms of throughput, latency, CPU and memory consumption.

\subsection{Results}
\label{subsec:res}
\vspace{2pt}
    \subsubsection{Throughput and latency for blockchain operations}
    \vspace{2pt}
    In this experiment, we examined the throughput and latency of different DIMY blockchain transactions of uploading CBF, token issue and querying through QBF, when a load of 50 tx/second was sent to the blockchain. We define throughput as the rate at which transactions are successfully executed and latency as the time required to complete the transactions. Please note that although a complete processing cycle of our architecture includes the serial execution of multiple DIMY transactions, we examined the throughput and latency only for each individual transaction. For instance, we assume that the CBF is already uploaded to the blockchain and we only measure the throughput and latency for executing query QBF. This definition does not consider the network latency, which could be impacted by different external factors. 
    
    We plot the results in Figure~\ref{fig:task-compare}. Among three DIMY blockchain transactions, QBF upload and matching is the transaction with the lowest latency, while uploading CBF has the highest latency, at around 4700 ms. Note that although we set a constant transaction send rate of 50 tx/second, issue tokens and upload CBF transactions failed to deliver the same throughput rate. The latency for uploading 50 simultaneous CBFs is the highest  when compared with other operations, as this involves transaction insertions and consensus operations at the blockchain. QBF matching, on the other hand, performs very well in terms of latency and throughput. We note that uploading CBF is only performed once for each identified COVID-19 positive case, while QBF upload and matching is performed once in a 24-hour cycle by each user. 
    
    %Despite the high latency identified in the upload CBF operation, the measured time would still be acceptable, because this scenario assumes that the health authority issues multiple tokens, and everyone is uploading their CBF simultaneously.
    
% \textcolor{blue}{For issue token, the latency is very low (about 175ms) why the throughput is affected, why it has dropped from 50 tx/sec?
% }
%\textcolor{blue}{Latency: does it include the complete processing cycle e.g., for QBF- upload QBF search for matching with existing transactions and return of results? or simply upload of QBF to the chain?}

    \subsubsection{CPU and memory consumption for blockchain operations}
    We compare the CPU and memory consumption of different operations and show the results in Figure~\ref{fig:cpu-mem}. A Caliper monitor was utilised to capture CPU and memory usage when we applied a load of 50 tx/second to the Hyperledger Fabric network for 30 seconds. The results show that although there seems to be no significant difference in average memory consumption, QBF matching consumes the highest CPU percentage (about 80\%), when compared with other operations. This result highlights that to host the backend blockchain, the back-end's memory and CPU requirements are the key design factors, as QBF matching is the most used operation on the backend.

    \subsubsection{Latency of querying QBF with different sized Bloom filters}
    The following experiment aims to examine the effect of using different sized Bloom filters on the resulting latency of QBF matching operations on the blockchain. We executed different transaction rate loads, namely 75, 100 and 125 tx/second,  noting both maximum and the average latency for QBF sizes varied from 10KB to 100KB. The results shown in Figure~\ref{fig:varying-qbf} demonstrate that the average latency remains less than 100 ms for all QBF sizes less than 900KB, while the maximum observed latency remains less than 100 ms for QBF sizes up to 100KB. The maximum latency starts increasing considerably, especially for higher transaction rates, if the size of the QBF is increased beyond 100KB. This result shows that our chosen value of 100KB for Bloom filters is optimal to minimise the maximum observed latency for different transaction rates.  
    
    %The average latency for all different send rate remains stable at a rate less than 100 ms, although there is a slight increase for 125 tx/second rate when Bloom filter size exceeds 800KB. For all transaction send rates, there is a similar trend in which latency increases gradually as the size of the Bloom filter increases.
% \textcolor{blue}{Difficult to comprehend this experiment as it does not fit well with our proposed scheme where we use the same size 100KB for all BFs (DBF, CBF and QBF). We could extend this graph with lower values of bloom filter size less than 100KB and then perhaps comment on the efficacy of choosing an appropriate size of the BF wrt latency}  

    \subsubsection{Throughput and latency analysis}
    Lastly, we investigate the throughput and latencies for querying QBF with different number of Hyperledger peer nodes (2 , 4 and 8 nodes) and plot the results in Figure~\ref{fig:throuhgput}. During this experiment, we started from 50 QBF tx/second and gradually increased the send rate up to 500 tx/second. In this experiment, we noticed that the blockchain was still able to deliver increasing throughput in line with the transaction send rate. % that in Figure~\ref{fig:task-compare}, issue tokens and upload CBF transactions fail to achieve a similar throughput as the transaction send rate. 
    Besides, maximum latency remains below 100 ms for transaction rate of up to 250 tx/second for all explored cases, after which there is an observed increase for 2 nodes Hyperledger. However, the average latency stays low (less than 50 ms) for all cases.
    We note that this observed performance is achieved with the hardware resources used for this proof of concept described in Section \ref{subsec:poc}, which can be easily scaled up in the actual deployment of the proposed architecture.
\vspace{-5pt}

\section{Conclusion}
\label{sec:Conc}
In this paper, we have presented the design and security and privacy evaluation for DIMY, a privacy-preserving digital contact tracing protocol. Our protocol design integrates several privacy preserving techniques, assuming both malicious users and the back-end as  the threat model. We employed a Bloom filter to enhance privacy protection as well as to considerably cut down storage requirements both on the client's device and the back-end.

Our protocol is resilient against most of the security and privacy attacks commonly launched against digital contact tracing apps. The proposed protocol incurs negligible overheads and supports low latency operations on the backend side, as demonstrated in our performance evaluations. The development of the open-source app and its GUI is part of our future work.%We are currently working on the open-source prototype of the mobile app with the DIMY design.
\vspace{-5pt}

% or


%\appendices

%\section{}

\ifCLASSOPTIONcompsoc
  \section*{Acknowledgments}
\else
  \section*{Acknowledgment}
\fi

This work has been supported by the Cyber Security Cooperative Research Centre Limited (CSCRC), whose activities are partially funded by the Australian Government’s Cooperative Research Centres Programme. 
%\vspace{-5pt}

\ifCLASSOPTIONcaptionsoff
  \newpage
\fi



\bibliographystyle{IEEEtran}
\bibliography{references}
\vspace{-40pt}
\begin{IEEEbiography}[{\includegraphics[width=1in,height=1.25in,clip,keepaspectratio]{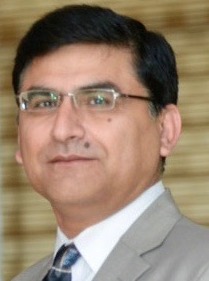}}]{Nadeem Ahmed} received M.S. and Ph.D. degrees in computer science from the UNSW, Sydney, Australia, in 2000 and 2007, respectively. He is currently working as a senior research fellow at the Cyber Security Cooperative Research Centre (CSCRC), Australia.  Earlier, he worked as head of the Computing Department at the School of Electrical Engineering and Computer Science, NUST, Pakistan. His research interests include cyber security, IoT, wireless sensor networks, and software-defined networking.
\end{IEEEbiography}
\vspace{-40pt}
\begin{IEEEbiography}[{\includegraphics[width=1in,height=1.25in,clip,keepaspectratio]{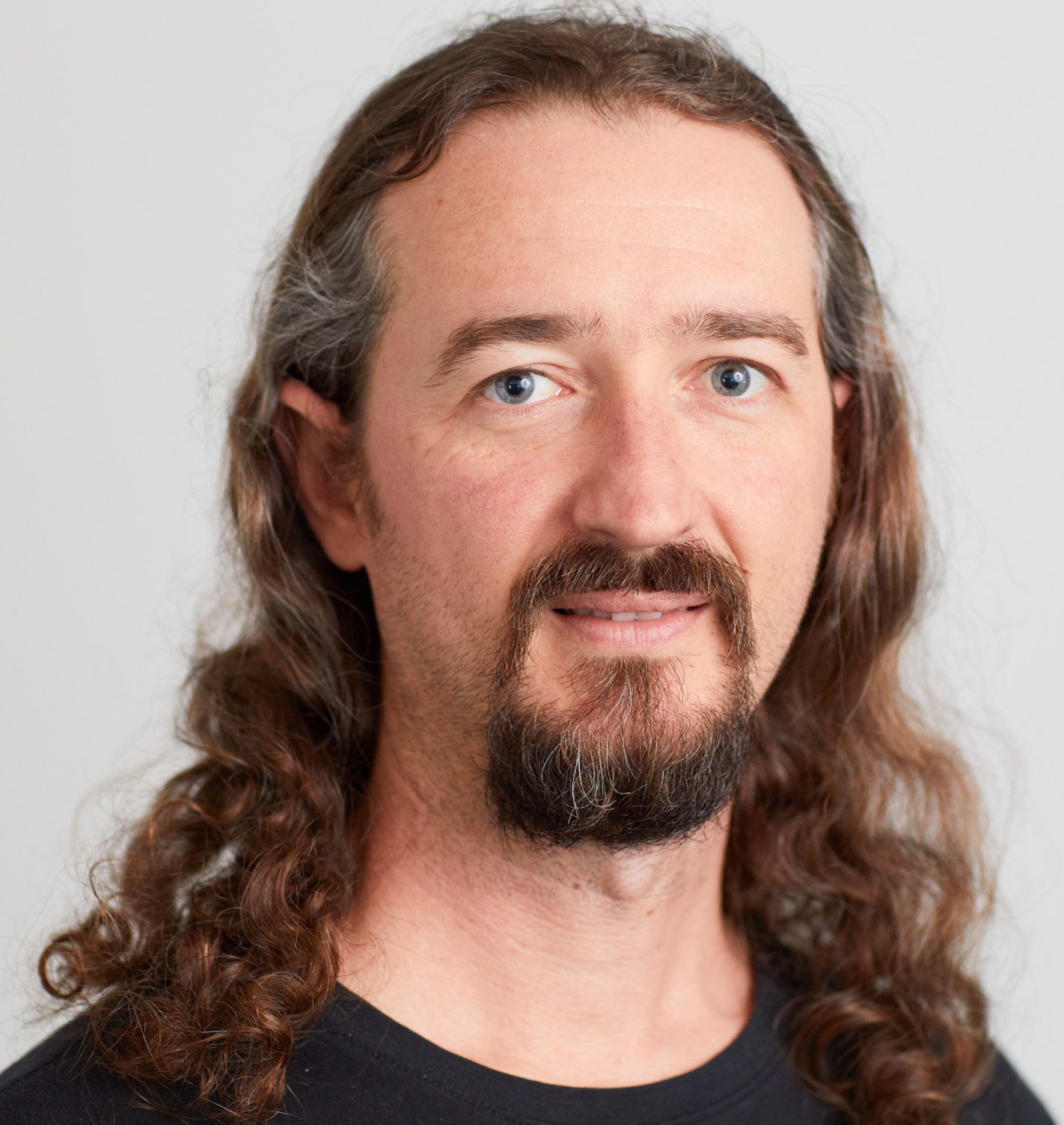}}]{Regio A. Michelin} received M.S. and Ph.D. degrees in computer science from the Pontifical Catholic University of Rio Grande do Sul, Brazil, in 2014 and 2019, respectively. He is currently working as research fellow at the Cyber Security Cooperative Research Centre (CSCRC), Australia. His research interests include blockchain, cybersecurity, and IoT.
\end{IEEEbiography}
\vspace{-40pt}
\begin{IEEEbiography}[{\includegraphics[width=1in,height=1.25in,clip,keepaspectratio]{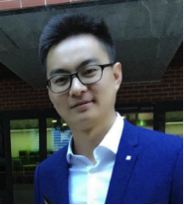}}]{Wanli Xue} received Ph.D.
degree from the School of Computer Science
and Engineering, UNSW, Australia. He is currently a Research Fellow at the Cyber Security Cooperative Research Centre (CSCRC) and UNSW, Australia. His research interests include security and
privacy issues in cyber physical systems and IoT,
including highly efficient privacy-preserving techniques for IoT as well as IoT-related sensing systems
and data analytic services.
\end{IEEEbiography}
\vspace{-40pt}
\begin{IEEEbiography}[{\includegraphics[width=1in,height=1.25in,clip,keepaspectratio]{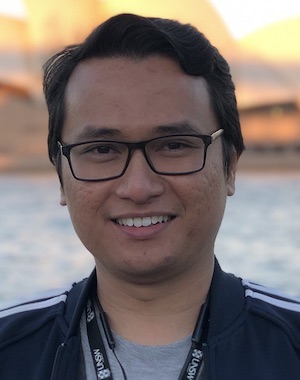}}]{Guntur Dharma Putra} received his bachelor degree in Electrical Engineering from Universitas Gadjah Mada, Indonesia, in 2014. He received his master's degree in Computing Science from the University of Groningen, the Netherlands, in 2017. He is currently a Ph.D. candidate at UNSW, Sydney, Australia. His research interests cover distributed systems and the IoT. He also looks into blockchain applications for securing IoT. Guntur is a student member of the IEEE.
\end{IEEEbiography}
\vspace{-40pt}
\begin{IEEEbiography}[{\includegraphics[width=1in,height=1.25in,clip,keepaspectratio]{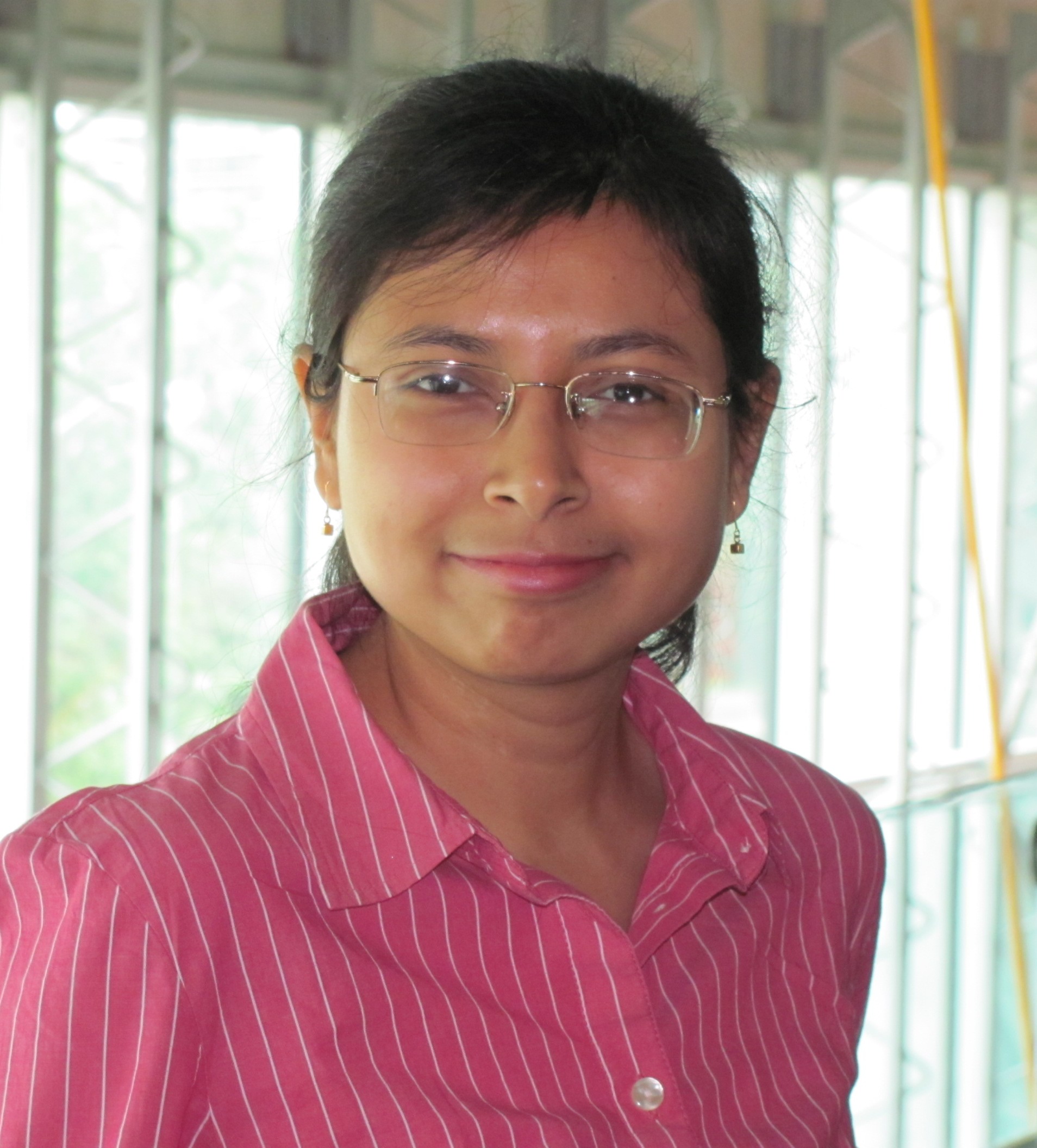}}]{Sushmita Ruj} received her Masters and PhD in Computer Science from the Indian Statistical Institute. She is currently a Senior Research Scientist at CSIRO Data61, Australia. She is also an Associate Professor at Indian Statistical Institute, Kolkata. Her research interests include blockchains, applied cryptography, and data privacy. She serves as a reviewer of Mathematical Reviews, and an Associate Editor of Elsevier Journal, Information Security and Applications. She is a recipient of the Samsung GRO award, NetApp Faculty Fellowship, Cisco Academic Grant and IBM OCSP grant. She is a Senior Member of the ACM and IEEE.
\end{IEEEbiography}
\vspace{-40pt}
\begin{IEEEbiography}[{\includegraphics[width=1in,height=1.25in,clip,keepaspectratio]{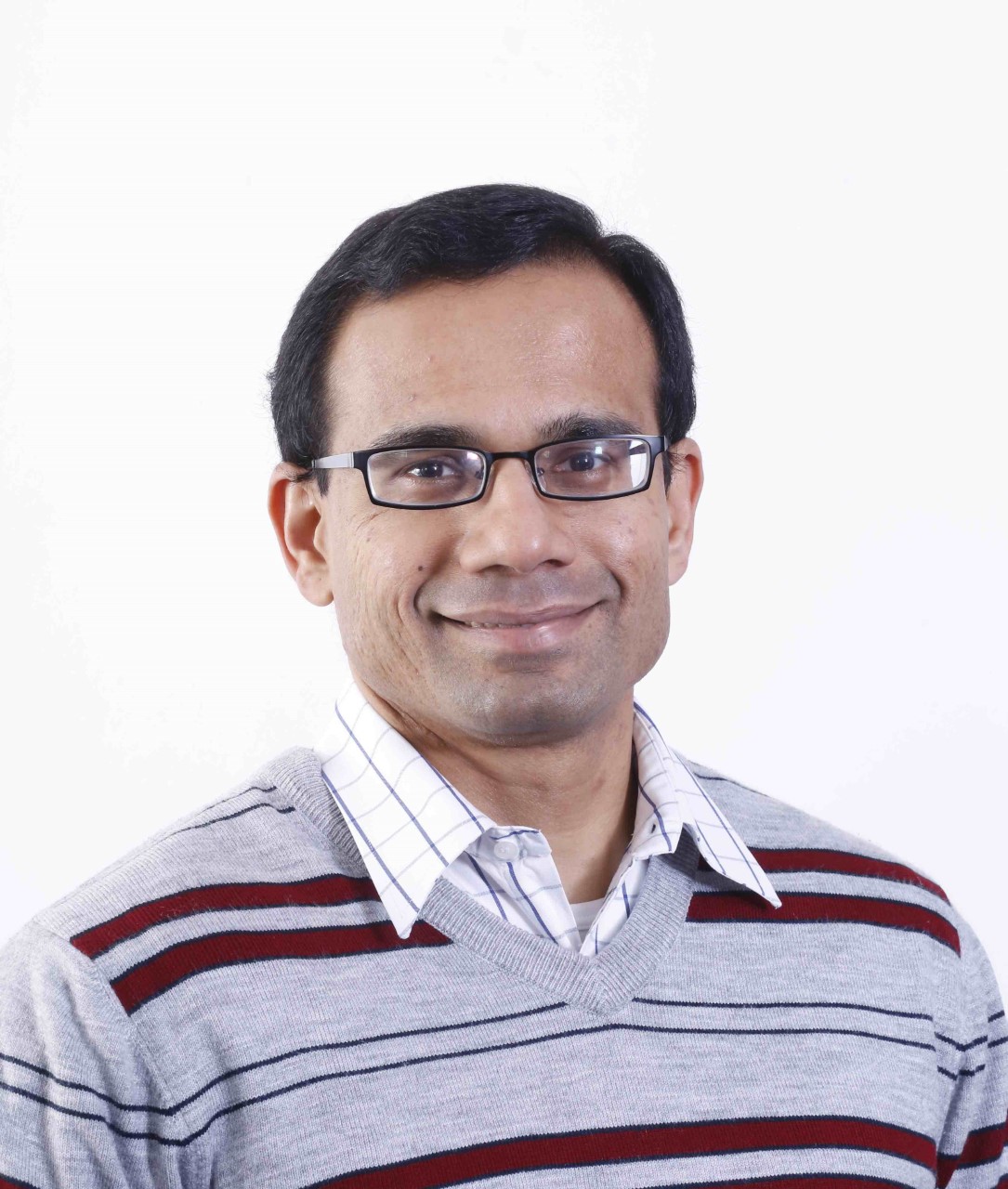}}]{Salil S. Kanhere} received his M.S. and Ph.D. degrees from Drexel University in Philadelphia. He is a Professor of Computer Science and Engineering at UNSW Sydney, Australia. His research interests include the IoT, blockchain, pervasive computing, cybersecurity and applied machine learning. He is a Senior Member of the IEEE and ACM, an Humboldt Research Fellow and an ACM Distinguished Speaker. He serves as the Editor in Chief of the Ad Hoc Networks journal and as Associate Editor of IEEE TNSM, COMCOM and PMC. He has served on the organising committee of several IEEE/ACM international conferences.
\end{IEEEbiography}
\vspace{-40pt}
\begin{IEEEbiography}[{\includegraphics[width=1in,height=1.25in,clip,keepaspectratio]{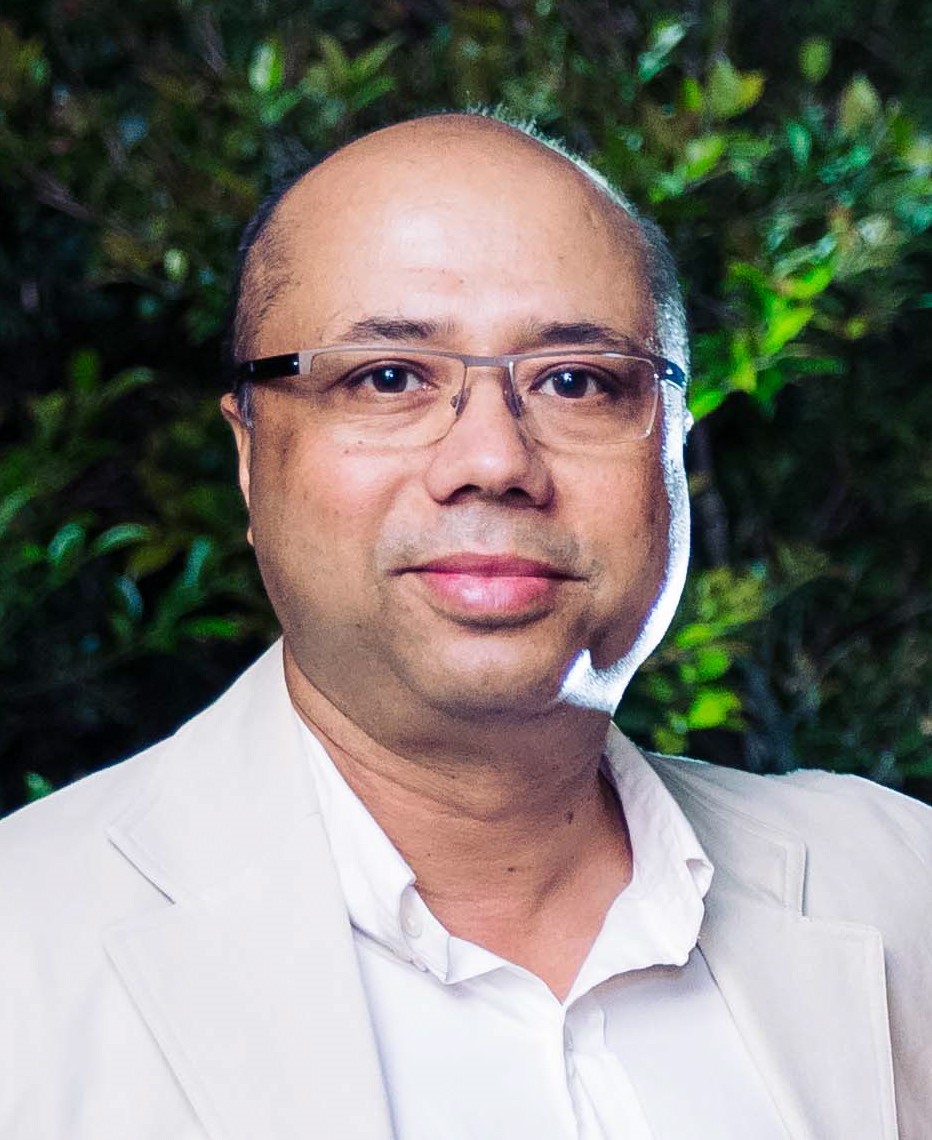}}]{Sanjay K. Jha} is a Full Professor and Director of the Cybersecurity and Privacy Lab at the School of Computer Science and Engineering at the UNSW, Australia. He leads UNSW in the Cybersecurity Cooperative Research Centre. His research activities are primarily focused on Wireless Mesh/Sensor Networks (IoT), and Network Security.  He is the principal author of the book "Engineering Internet QoS" and a co-editor of the book "Wireless Sensor Networks: A Systems Perspective." His  editorial affiliations include the IEEE TMC and TDSC.
\end{IEEEbiography}


\end{document}